\documentclass[aps,prl,twocolumn,nofootinbib,superscriptaddress,showpacs,floatfix,preprintnumbers]{revtex4-2}

\usepackage[dvipsnames]{xcolor}     
\definecolor{lcolor}{rgb}{0.5,0,0}
\definecolor{citcolor}{rgb}{0,0,1}
\usepackage[breaklinks,colorlinks,urlcolor=blue,citecolor=citcolor,linkcolor=lcolor,linktoc=all]{hyperref}
\usepackage{color}
\usepackage{graphicx}	
\graphicspath{{./figures/}}
\usepackage[utf8]{inputenc}
\usepackage{amsmath,euscript,array,mathrsfs,bigints}
\usepackage{amssymb}
\usepackage[dvipsnames]{xcolor}      
\usepackage{bm,bbm,bbold}
\usepackage{booktabs}
\usepackage{dcolumn}

\makeatletter
\def\@fnsymbol#1{\ensuremath{\ifcase#1\or *\or \dagger\or \ddagger\or
   \mathsection\or \mathparagraph\or \|\or **\or \dagger\dagger
   \or \ddagger\ddagger \else\@ctrerr\fi}}
\makeatother

\allowdisplaybreaks

\makeatletter
\g@addto@macro\bfseries{\boldmath}
\makeatother

\usepackage{tikz}
\usepackage[customcolors]{hf-tikz}
\usepackage{mciteplus}

\makeatletter 

\renewcommand\onecolumngrid{% <<<<<<
\do@columngrid{one}{\@ne}%
\def\set@footnotewidth{\onecolumngrid}% <<<<<<<<<<<<<<<<
\def\footnoterule{\kern-6pt\hrule width 1.5in\kern6pt}%
}

\renewcommand\twocolumngrid{% <<<<<<
        \def\footnoterule{% restore rule
        \dimen@\skip\footins\divide\dimen@\thr@@
        \kern-\dimen@\hrule width.5in\kern\dimen@}
        \do@columngrid{mlt}{\tw@}
}%

\makeatother    
\newcommand{\dd}{\mathrm{d}}

\usepackage[T1]{fontenc} 
\usepackage{graphicx}
\usepackage{epsfig}
\usepackage{rotating}
\usepackage{dsfont}
\usepackage{psfrag}
\usepackage{bbold}
\usepackage{epsf}
\usepackage{slashed}
\usepackage{cancel}
\usepackage{float}
\usepackage{comment}
\usepackage{caption,subcaption,wrapfig}
  \newcommand{\cJ}{{\cal J}}
\newcommand{\cK}{{\cal K}}

\newcommand{\cS}{{\cal S}}

\def\be{\begin{equation}}
\def\ee{\end{equation}}
\def\ba{\begin{eqnarray}}
\def\ea{\end{eqnarray}}

\def\dd{\mathrm{d}}
\def\DD{\mathrm{D}}

\newcommand{\corr}[1]{\langle#1\rangle}

\usepackage{pifont}% More styles for bullets

\begin{document}  

\title{Thermal response of the Nieh--Yan term}

\preprint{HIP-2024-26/TH}

\author{Carlos Hoyos}
\email{hoyoscarlos@uniovi.es}
\affiliation{Departamento de F\'{\i}sica and Instituto de Ciencias y Tecnolog\'{\i}as Espaciales de Asturias (ICTEA), Universidad de Oviedo,c/ Leopoldo Calvo Sotelo 18, ES-33007, Oviedo, Spain}
\author{Niko Jokela}
\email{niko.jokela@helsinki.fi}
\affiliation{Department of Physics and Helsinki Institute of Physics,
P.O.~Box 64, FI-00014 University of Helsinki, Finland}
\author{Jos\'e Manuel Pen\'in}
\email{jmanpen@gmail.com}
\affiliation{INFN, Sezione di Firene; Via G. Sansone 1; I-50019 Sesto Fiorentino (Firenze), Italy }
\affiliation{Dipartimento di Fisica e Astronomia, Universit\'a di Firenze, \\ Via G. Sansone 1; I-50019 Sesto Fiorentino (Firenze), Italy }

\begin{abstract}
%We reinterpret the Nieh–-Yan anomaly using holography, finding that the $U(1)$ axial symmetry remains unbroken and the axial current coupling to an external gauge field is conserved. Instead, the anomaly arises from a breakdown of Hodge duality relations between fermion bilinears due to symmetry constraints on one-form currents. We also compute the thermal response of the Nieh–-Yan term, revealing its characteristic $T^2$ dependence.
We reinterpret the Nieh-Yan (NY) anomaly using holography, finding that the U(1) axial symmetry remains unbroken and the axial current coupling to an external gauge field is conserved. Instead, the anomaly arises from a breakdown of Hodge duality relations between fermion bilinears due to symmetry constraints on one-form currents. 
%Using a holographic model where axial anomalies are absent, 
We show that the axial response associated with the NY term is distinct from the chiral vortical effect and exhibits a cha\-rac\-te\-ris\-tic $T^2$ dependence. Torsion-induced axial transport decouples spin and axial charge dynamics, thereby clarifying the physical significance of the NY anomaly and motivating further field-theoretic and holographic studies.
\end{abstract}

\maketitle

\section{Introduction}

Chirality is a fundamental property of relativistic fermions, prominently featured in high energy physics and the effective theories of Weyl semimetals \cite{Yan_2017} and time-reversal-breaking superfluid and superconductor phases, such as ${}^3$He-A \cite{Volovik:2003fe,Meng_2012}. A notable consequence is the emergence of unusual charge and thermal transport phenomena in response to external sources. This is often associated with anomalies in the axial current, which measure the difference between currents with opposite chiralities \cite{Chernodub:2021nff}. Anomalous transport may leave observable imprints on the charge distribution of particles produced in heavy-ion collisions \cite{Fukushima:2008xe,Fukushima:2010vw,Kharzeev:2013ffa}, an effect currently under investigation \cite{STAR:2009wot,Li:2014bha,Kharzeev:2024zzm}. Axial anomalies have also been studied in Weyl semimetals \cite{Nielsen:1983rb,Son_2013,Lundgren:2014hra,Burkov_2015,Spivak_2016,Lucas:2016omy}, and have been invoked to explain the large negative magnetoresistance measured in experiments \cite{Kim_2013,doi:10.1126/science.aac6089,Huang_2015,Li_2015,Li_2016,Zhang_2016,Hirschberger_2016,Gooth:2017mbd}.

An intriguing extension of these phenomena involves torsion-induced transport. Although torsion appears to be absent in the fabric of spacetime, it emerges effectively in condensed matter systems as lattice defects in solids \cite{deJuan:2009ldt,Hughes:2012vg, Parrikar:2014usa} or textures of the order parameter in superfluids or superconductors \cite{Ishihara_2019}. 
The axial response to torsion in condensed matter systems has become a topic of significant interest, yet it remains highly debated and not fully understood.

Although the anomalous axial response to external gauge fields and background geometry is well established, the response to torsion remains puzzling. The putative torsional anomaly of the axial current, $\cJ_{1,5}^\mu = \bar{\psi} \gamma^\mu \gamma_5 \psi$, is characterized by a topological term introduced by Nieh and Yan (NY) \cite{Nieh:1981ww}: 
\begin{equation}\label{eq:NYterm} 
\partial_\mu \corr{\cJ_{1,5}^\mu}=-ic_\textrm{NY}\epsilon^{\mu\nu\lambda\rho}\left( \eta_{\alpha\beta} T^\alpha_{\mu\nu} T^\beta_{\lambda\rho}-\frac{1}{2}R_{\mu\nu\lambda\rho}\right) \, , 
\end{equation} 
where $T^\alpha_{\mu\nu}$ and $R_{\mu\nu\lambda\rho}$ denote the torsion and Riemann curvature tensors, respectively. Unlike other anomalies, the coefficient $c_\textrm{NY}$ is dimensionful. Due to the absence of any natural scale, its presence in high energy physics is dubious \cite{Chandia:1997hu,Obukhov:1997pz, Kreimer:1999yp,Chandia:1999az,Erdmenger:2024zty}. However, in condensed matter systems $c_\textrm{NY}$ can acquire a natural interpretation \cite{Parrikar:2014usa}, being proportional to a cutoff of the effective theory, to temperature $T$, or to various chemical potentials~\cite{Sumiyoshi:2015eda,Ferreiros:2018udw, Huang_2019, Huang:2020ypv, Huang_2020b, Nissinen:2019kld, Laurila:2020yll}. Furthermore, a universal temperature-dependent contribution, $c_{\textrm{NY}} \propto T^2$, has been proposed \cite{Nissinen:2019wmh, Nissinen:2019mkw, Huang_2020, Liang_2020, Khaidukov:2018oat, Imaki:2019ite, Imaki:2020csc,Nissinen:2021gke}. 
%However, there is an ongoing debate \cite{Chernodub:2021nff, Ferreiros:2020uda, Amitani:2022xev, Valle:2021nfv, Valle:2023cqo} about whether this represents a genuine axial-torsional effect or is merely the chiral vortical effect \cite{Landsteiner:2011iq, Braguta:2013loa} in disguise.
The nature of this response remains disputed \cite{Chernodub:2021nff,Ferreiros:2020uda,Amitani:2022xev,Valle:2021nfv,Valle:2023cqo}, with some questioning whether it genuinely reflects an axial-torsional effect or merely disguises the chiral vortical effect \cite{Landsteiner:2011iq,Braguta:2013loa}.

%In this work, we argue that the thermal axial response arising from a NY term is indeed possible and distinct from the chiral vortical effect, even though it does not constitute an anomaly in the conventional sense. 
%Our central observation is that, while previous arguments are correct regarding the response in the consistent axial current, the complete dismissal of the torsional response relies on identifying the bilinear coupling of the fermion to torsion, $\cJ_3^{\mu\nu\lambda} = \bar{\psi} \gamma^{[\mu} \gamma^\nu \gamma^{\lambda]} \psi$, with the Hodge dual of the axial current. 
In this Letter, we demonstrate that the thermal axial response arising from (\ref{eq:NYterm}) is indeed distinct from the chiral vortical effect, although it does not constitute an anomaly in the conventional sense. The key insight is that previous arguments correctly address the response in the consistent axial current, but the outright dismissal of the torsional response depends on an oversimplified identification of the bilinear coupling of the fermion to torsion,
$\cJ_3^{\mu\nu\lambda} = \bar{\psi} \gamma^{[\mu} \gamma^\nu \gamma^{\lambda]} \psi$,
with the Hodge dual of the axial current. 
However, in the absence of anomalies, the quantum expectation value $\corr{\cJ_{1,5}^\mu}$ represents a conserved current with only transverse components, while $\corr{\cJ_3^{\mu\nu\lambda}}$ may have both transverse and longitudinal components. In the presence of torsion $\corr{\cJ_3^{\mu\nu\lambda}}$ can develop a transverse component, such that its Hodge dual no longer coincides with the conserved axial current but instead corresponds to an axial vector operator with a non-conservation law compatible with the NY anomaly. Intriguingly, this decouples spin and axial charge dynamics.

%\tb{The argument presented above is completely general, but to illustrate our proposal in practice, we compute the relevant expectation values in a holographic model with axial anomalies turned off.} 
%This requires careful implementation of the Hodge dual relations among fermion bilinears, which we systematically verify in all cases. Our results agree with the interpretation of the NY anomaly given above. \tb{We encourage a direct verification of our proposal through a field theory calculation,
%which could be done by a direct calculation of the expectation values of the fermion bilinears in the presence of a background torsion using standard quantum field theory tools, or by revisiting previous  works \cite{Sumiyoshi:2015eda,Ferreiros:2018udw, Huang_2019, Huang:2020ypv, Huang_2020b, Nissinen:2019kld, Laurila:2020yll,Nissinen:2019wmh, Nissinen:2019mkw, Huang_2020, Liang_2020, Khaidukov:2018oat, Imaki:2019ite, Imaki:2020csc,Nissinen:2021gke} and carefully identifying torsional sources and consistent currents.
%In addition, the model we present provides a novel realization of spin dynamics in holography, which has been studied in \cite{Leigh:2008tt,Petkou:2010ve,Gallegos:2020otk,Erdmenger:2022nhz,Erdmenger:2023hne,Cartwright:2024dcj} partially motivated by the hadron polarization observed in heavy-ion collisions \cite{STAR:2017ckg,STAR:2019erd,ALICE:2019aid,STAR:2020xbm}. Since in our formulation spin and chiral dynamics can be disentangled, there could be spin polarization effects in the rotating plasma independent of vortical effects in the axial current.}

To illustrate our proposal in a controlled setting, we compute the relevant expectation values in a holographic model where axial anomalies are turned off, definitely removing possible contributions from a chiral vortical effect. A proper implementation of torsional response necessitates a meticulous implementation of Hodge duality relations among fermion bilinears in the holographic dual, which we systematically verify. Our results align with the proposed interpretation of the NY anomaly, and we advocate for a direct verification through field-theoretic calculations. Such an analysis could involve explicit computations of fermion bilinear expectation values in a background torsion field using standard quantum field theory tools or a careful re-examination of prior works \cite{Sumiyoshi:2015eda,Ferreiros:2018udw,Huang_2019,Huang:2020ypv,Huang_2020b,Nissinen:2019kld,Laurila:2020yll,Nissinen:2019wmh,Nissinen:2019mkw,Huang_2020,Liang_2020,Khaidukov:2018oat,Imaki:2019ite,Imaki:2020csc,Nissinen:2021gke} to precisely identify torsional sources and consistent currents.

Furthermore, our holographic model offers a novel perspective on spin dynamics in strongly coupled systems, an area of active research \cite{Leigh:2008tt,Petkou:2010ve,Gallegos:2020otk,Gallegos:2022jow,Erdmenger:2022nhz,Erdmenger:2023hne,Cartwright:2024dcj}. This is particularly relevant in the context of hadron polarization observed in heavy-ion collisions \cite{STAR:2017ckg,STAR:2019erd,ALICE:2019aid,STAR:2020xbm}, as our approach enables a clear distinction between spin polarization and effects due to chiral imbalance in a rotating plasma.

Our findings reveal a distinct and physically meaningful axial response to torsion in condensed matter systems, offering new insights into the interplay between torsion, axial anomalies, and spin dynamics. Future investigations, both in field theory and holography, could further elucidate the significance of the NY anomaly and its experimental manifestations.

\section{Duality relations for fermion bilinears}

Due to the properties of the Clifford algebra in four dimensions, not all tensor operators constructed as fermion bilinears are independent, but are related by a Hodge duality.  For a Dirac fermion $\psi$ we can construct completely antisymmetric tensor operators
\begin{subequations}
\begin{eqnarray}
    \cJ_p^{\mu_1\cdots \mu_p} &= & \overline{\psi}\gamma^{[\mu_1}\cdots \gamma^{\mu_p]}\psi \\ \cJ_{p,5}^{\mu_1\cdots \mu_p} &=& \overline{\psi}\gamma^{[\mu_1}\cdots \gamma^{\mu_p]}\gamma_5\psi \ ,
\end{eqnarray}
\end{subequations}
where $\gamma_5=i\gamma^0 \gamma^1 \gamma^2 \gamma^3$ and we have grouped the operators transforming either vectorially or axially. Here, $p=0$ corresponds to scalar and pseudoscalar operators. These operators can be identified as components of $p$-forms
\begin{equation}
    \cJ_{p(,5)}=\frac{1}{p!}\cJ_{p(,5) \,\mu_1\cdots \mu_p} \dd x^{\mu_1}\wedge \cdots \wedge \dd x^{\mu_p} \ .
\end{equation}
It is convenient to introduce complex combinations 
\begin{equation}
    \cJ_p^\pm=\frac{1}{2}\left(\cJ_p\pm i \cJ_{p,5}\right) \ .
\end{equation}
Denoting by $\star_4$ the Hodge dual operation in four dimensions, the relations between operators derived from the Clifford algebra are
\begin{equation}\label{eq:vevduality}
\cJ_4^\pm=\pm \star_4 \cJ_0^\mp\ , \quad \cJ_3^\pm=\pm \star_4 \cJ_1^\mp\ ,\quad \cJ_2^\pm=\mp\star_4 \cJ_2^\mp \ .
\end{equation}
One can introduce external $p$-forms $c_p^\pm$ acting as sources for each of these operators
\begin{equation}\label{eq:ap}
    I_p=\int c_p^\pm\wedge \star_4 \cJ_p^\pm=-\int \dd^4x \frac{1}{p!}c_{\mu_1\cdots \mu_p}^\pm\cJ_{p}^{\pm\,\mu_1\cdots \mu_p} \ .
\end{equation}
The relations \eqref{eq:vevduality} lead to duality relations for the sources
\begin{equation}\label{eq:dualityforms}
    c_4^\pm=\mp \star_4 c_0^\mp\ ,\quad c_3^\pm=\pm \star_4 c_1^\mp\ ,\quad c_2^\pm=\pm\star_4 c_2^\mp \ .
\end{equation}
When fermions are coupled to a curved background it is convenient to use the Cartan formalism with a vielbein and spin connection. Assuming flat spacetime, a non-trivial spin connection implies a non-zero torsion, which is minimally coupled to the fermions through the term \cite{Manes:2020zdd,Ferreiros:2020uda,Erdmenger:2024zty}
\begin{equation}
  -\frac{1}{16}\,\,\omega_{\mu\alpha\beta}\overline{\psi}  \{ \gamma^\mu,[\gamma^\alpha,\gamma^\beta]\}\, \psi \ .
\end{equation}
Using the identity $\{ \gamma^\mu,[\gamma^\alpha,\gamma^\beta]\}=4 \gamma^{[\mu}\gamma^\alpha\gamma^{\beta]}$, this contribution is equivalent to introducing three-form sources into the action
\begin{equation}\label{eq:spinconsources}
    c_{3\,\mu\nu\lambda}^+=c_{3\,\mu\nu\lambda}^-= \frac{3}{2} \omega_{[\mu\nu\lambda]}\ . 
\end{equation}
Since torsion and three-form sources are equivalent, the three-form operator is proportional to the completely antisymmetric part of the spin current $\corr{\cJ_3}=\frac{2}{3}\corr{\cS_3}$.

%%%%%%%%%%%%%%%%%%%%%%%%%%%%%%%%%%%%%%%%%%%
\section{Holographic dual}
%%%%%%%%%%%%%%%%%%%%%%%%%%%%%%%%%%%%%%%%%%%

According to the holographic dictionary, each bilinear fermion operator corresponds to a complex form field in the five-dimensional dual geometry
\begin{equation}
    \cJ_p^\pm\ \longleftrightarrow C_p^\pm \ .
\end{equation}
We will assume that the dual geometry is asymptotically $AdS_5$. The fermion bilinears are all of scaling dimension $\Delta=3$,  this fixes the masses of the $p$-forms in units of the $AdS_5$ radius $L$ of the holographic dual
\begin{equation}\label{eq:formmass}
    M_p^2L^2=(3-p)(p-1) \ .
\end{equation}

%%%%%%%%%%%%%%%%%%%%%%%%%%%%%%%%%%%%%%%%%%%%
\subsection{Action for even forms}
%%%%%%%%%%%%%%%%%%%%%%%%%%%%%%%%%%%%%%%%%%%%

We will start with the even forms and study the more complicated case of the odd forms separately. The five-dimensional action for the even forms is ($\DD C_p^\pm=\dd C_p^\pm- i(C_1^+-C_1^-) \wedge C_p^\pm$)
\begin{equation}
\begin{aligned}
\label{eq:S0S2}
&S=\sum_{p=0,2}\frac{\lambda_p}{2} \int \epsilon_{AB} (C_p^A \wedge \DD C^B_{4-p} -\DD C_p^A\wedge C^B_{4-p})\ \\&-\frac{\delta_{AB}}{1+\delta_{p,2}}(m_p\, C^{A}_p\,\wedge  \star C^{B}_p+m_{4-p} \, C^{A}_{4-p}\wedge \star C^{B}_{4-p}) \ ,
\end{aligned}
\end{equation}
where $A,B=\pm$, $\epsilon_{+-}=1$ and we have introduced the Hodge star operator in five dimensions $\star$ ($\epsilon^{01234}=1$). The masses in the action have the values
\begin{equation}\label{eq:m402}
    m_4L=1,\quad m_0L=-3,\quad m_2L=2\ .
\end{equation}
A non-Abelian version of the action for two-forms was originally introduced in \cite{Alvares:2011wb}.

%%%%%%%%%%%%%%%%%%%%%%%%%%%%%%%%%%%%%%%%%%%%
\subsection{Action for odd forms}
%%%%%%%%%%%%%%%%%%%%%%%%%%%%%%%%%%%%%%%%%%%%

Omitting the gravitational sector, the action for the odd forms is similar to the one for the even forms
\begin{equation}\label{eq:S1}
S_1= \lambda_1\bigintssss \epsilon_{AB} \dd C_1^A\wedge C^B_3-\frac{m_3}{2}\delta_{AB} C^{A}_3\wedge \star\, C^{B}_3.
\end{equation}
The action is invariant under a zero-form gauge symmetry
\begin{equation}\label{eq:gaugtr}
    \delta C_1^\pm=\dd \Lambda_0^\pm \ .
\end{equation}
This corresponds to the $U(1)\times U(1)$ global flavor symmetry in the dual field theory.

We now consider the coupling to the spin connection. In the Cartan formalism, the gravitational and torsional degrees of freedom are captured by the vielbeins $e^a$ and spin connection $\omega^a_{\ b}$ one-forms. The indices $\{a,b\}=\{0,1,2,3,z\}$ refer to the tangent space of the five-dimensional geometry. The torsion and curvature are given by the two-forms
\begin{align}
 T^a = \dd e^a +\omega^a_{\ b}\wedge e^b \ , \
    R^a_{\ b} = \dd\omega^a_{\ b}+\omega^a_{\ c}\wedge \omega^c_{\ b} \ ,
\end{align}
where indices are raised and lowered with the flat Minkowski metric.
The Einstein--Hilbert action with a cosmological constant $\Lambda$ is 
\begin{equation}\label{eq:Sgrav}
S_{\tiny \text{grav}}=\frac{1}{2\kappa^2}\int \bigg( \frac{1}{3!}{\cal R}-2\Lambda \star 1 \bigg)\ ,
\end{equation}
where
\begin{equation}
{\cal R}=    \epsilon_{abcde}e^{a}\wedge e^{b}\wedge e^{c}\wedge R^{de}
\end{equation}
and where $\kappa^2$ is the 5D gravitational constant.

The coupling to the odd forms will be through the Lorentz-invariant three-form
\begin{equation}
     \Omega_3=\eta_{ab} e^a\wedge T^b\ .
\end{equation}
It should be noted that in $3+1$ dimensions the NY term \eqref{eq:NYterm} equals $\star_4 \dd\Omega_3$. A minimal term preserving the zero-form  gauge symmetry \eqref{eq:gaugtr} is
\begin{equation}\label{eq:Stor}
\begin{aligned}
&S_{\tiny \text{tor}}=\frac{\lambda_t}{4\kappa^2}\int  \Omega_3 \wedge \left[m_3 \star (C_3^++C_3^-)- \dd (C_1^+-C_1^-) \right] \ .
\end{aligned}
\end{equation}
The last term of this action coincides with the one obtained in  \cite{Valle:2021nfv} using the descent formalism.

To fully determine our theory, we need to introduce additional terms in the action that describe the dynamics of torsion via the spin connection. The simplest action that can be constructed is as follows:
\begin{equation}\label{eq:Somega}
    S_\omega=\frac{L^2}{2\kappa^2 g_\omega^2}\int  R^a_{\ b}\wedge \star R^b_{\ a}+\frac{3}{2} \epsilon_{abcde}e^{a}\wedge R^{bc}\wedge R^{de}  \ .
\end{equation}
The values of the couplings that lead to a consistent system of equations are
\begin{equation}
    g_w^2=6\ , \ \Lambda L^2=-3\ , \  27 m_3 \kappa^2 \lambda_1 =1\ ,  \ \lambda_t=6\kappa^2 \lambda_1 \ .
\end{equation}
With this choice, the action contains the terms of the Lovelock--Chern--Simons action \cite{Gallegos:2020otk,Banados:2006fe,Erdmenger:2022nhz}.

%%%%%%%%%%%%%%%%%%%%%%%%%%%%%%%%%%%%%%%%%%%
\subsection{Equations of motion}
%%%%%%%%%%%%%%%%%%%%%%%%%%%%%%%%%%%%%%%%%%%

The equations of motion of the forms are obtained from the variation of \eqref{eq:S0S2}, \eqref{eq:S1}, and \eqref{eq:Stor}
\begin{equation}\label{eq:EQeven}
\dd \left(C^\pm_p-\frac{3}{2}\delta_{p,3}\Omega_3 \right)\pm \frac{m_{4-p}}{1+\delta_{p,2}} \star \left(C^\mp_{4-p}-\frac{3}{2}\delta_{p,1}\Omega_3 \right)=0\ .
\end{equation}
Combining the equations of motion we see that the forms satisfy the second-order equations
\begin{equation}\label{eq:Cpeqs}
   \star \dd \star \dd C_p^\pm +M_p^2 C_p^\pm=\frac{3}{2}\delta_{p,3}\star \dd\star \Omega_3  \ .
\end{equation}
The equations for the spin connection are derived from the variations of \eqref{eq:Sgrav}, \eqref{eq:Stor}, and \eqref{eq:Somega}. We decompose the spin connection in the background Levi--Civita connection $\mathring{\omega}$ plus a contorsion term which we treat as a fluctuation 
\begin{equation}
    \omega^a_{\ b}=\mathring{\omega}^a_{\ b}(e)+K^a_{\ b} \ .
\end{equation}
The linearized equations around the background $AdS_5$ geometry produce
\begin{equation}\label{eq:omegaeq}
     \mathring{D} \star \mathring{D} K^a_{\ b}+\frac{3}{2L^2} \epsilon^a_{\ b cde} K^{c}_{\ f} \wedge  e^f\wedge e^{d}\wedge e^{e}=0 \, ,
\end{equation}
where $\mathring{D}$ is the covariant derivative constructed with the background spin connection $\mathring{\omega}$.

%%%%%%%%%%%%%%%%%%%%%%%%%%%%%%%%%%%%%%%%%%%
\subsection{Asymptotic expansions}
%%%%%%%%%%%%%%%%%%%%%%%%%%%%%%%%%%%%%%%%%%%

We will consider an asymptotically $AdS_5$ black brane background
\begin{equation}\label{eq:AdS5met}
\dd s^2=\frac{L^2}{z^2}\left(\frac{\dd z^2}{f(z)}-f(z)\dd t^2+(\dd x^i)^2\right)\, ,\, f(z)=1-\frac{z^4}{z^4_h} 
\end{equation}
which, for vanishing background values of the forms, is a solution to the equations of motion. Here $z_h$ is the position of the horizon which is related to the temperature $T=(\pi z_h)^{-1}$. 
The asymptotic boundary is at $z=0$. 

The asymptotic form of the solutions to the second-order equations \eqref{eq:Cpeqs} is
\begin{equation}\label{eq:formexp}
    C_p^\pm \sim c_p^\pm \left(\frac{z}{L}\right)^{1-p}+k_p^{\pm} \left(\frac{z}{L}\right)^{3-p}\log \frac{z}{L}+j_p^\pm \left(\frac{z}{L}\right)^{3-p} \ .
\end{equation}
The factors $c_p^\pm$, $k_p^\pm$, and $j_p^\pm$ are $p$-forms along the field theory directions depending on the $x^\mu$ coordinates. $c_p^\pm$ and $j_p^\pm$ are respectively the coefficients of the `leading' and `subleading' solutions, while $k_p^\pm\sim \Box c_p^\pm$. We identify the components of $c_p^\pm$ along the field theory directions as sources for the corresponding dual bilinear operators, while the components of $j_p^\pm$ determine their vacuum expectation values.

 We can further simplify the equations of the spin connection \eqref{eq:omegaeq}. We will work in the gauge $K^a_{\ b z}=0$ and expand the contorsion in the vielbein basis
\begin{equation}
    K^a_{\ b}=\Omega^a_{\ bc}e^c, \ \ \Omega_3=\Omega_{[abc]} e^a \wedge e^b \wedge e^c \ .
\end{equation}
The non-zero components of the solutions are
\begin{equation}
    \Omega_{abc}=-\epsilon_{abcd}\eta^{d\mu} \partial_\mu \Omega_0\ , \ a,b,c,\mu\neq z\ .
\end{equation}
Here $\Omega_0$ is a scalar satisfying the same second-order equations as $C_0$ in the $AdS$ background. Therefore, the boundary expansion of the scalar coincides with \eqref{eq:formexp} for $p=0$, taking $c_0=\omega_0^{(0)}$, $j_0=S_0$. Then, the leading terms in the boundary expansion of the $\Omega_3$ components  coincide with \eqref{eq:formexp} for $p=3$, identifying
\begin{equation}\label{eq:spinconcoeff}
    \omega_3^{(0)} =\star_4 \dd\omega_0^{(0)},\qquad
    S_3= \star_4 \dd S_0 \ .
\end{equation}
This is the same kind of expansion we have for a massless three-form. Note that in the gauge we have chosen the coefficients of the asymptotic expansion are transverse,
\begin{equation}
    \partial^\mu \omega_{\mu\nu\lambda}^{(0)}=0=\partial^\mu S_{\mu\nu\lambda} \ .
\end{equation}

From the first-order equations \eqref{eq:EQeven} we derive additional conditions relating the coefficients of the asymptotic expansions that reproduce \eqref{eq:dualityforms} for the coefficients of the leading terms, except for $c_3^{\pm}$ which agrees with \eqref{eq:spinconsources}. There are similar relations for the subleading coefficients (see Supplemental material for the explicit expressions), in particular for orthogonal components
\begin{equation}\label{eq:j3coef}
    j_3^\pm=\frac{3}{2}S_3^\perp\pm\frac{1}{m_3 L}\star_4\left(2j_1^{\perp\mp}-\frac{L^2}{2}\Box c_1^{\perp\mp}\right) \ .
\end{equation}
The longitudinal components vanish $j_1^\parallel=S_3^\parallel=0$.

%%%%%%%%%%%%%%%%%%%%%%%%%%%%%%%%%%%%%%%%%%%
\section{Expectation values from holography}
%%%%%%%%%%%%%%%%%%%%%%%%%%%%%%%%%%%%%%%%%%%

According to the holographic dictionary, the generating functional of the quantum field theory equals the on-shell action of the gravity dual~\cite{Gubser:1998bc,Witten:1998qj}. Correlation functions of fermion bilinears are then obtained from the variation of the on-shell action with respect to the leading coefficients in the expansion of the forms \eqref{eq:formexp}. It is important to note that the currents obtained in this way are always {\em consistent} currents. 

The action is usually divergent and needs to be regularized and the divergences removed from physical quantities. This is done following the procedure of holographic renormalization \cite{deHaro:2000vlm,Bianchi:2001kw}, by which we introduce a cutoff in the holographic radial direction and add a counterterm action to remove the divergences.

%%%%%%%%%%%%%%%%%%%%%%%%%%%%%%%%%%%%%%%%%%%
\subsection{Variation of the action}
%%%%%%%%%%%%%%%%%%%%%%%%%%%%%%%%%%%%%%%%%%%

The variation of the five-dimensional on-shell action with respect to each form is a total derivative term
\begin{equation}\label{eq:deltaSon}
    \delta S_\textrm{on-shell} =-\sum_{p=0}^4 \sigma_p\int \dd\left[\delta_{AB}\delta C_p^A \wedge \star \dd C_p^B \right] \ ,
\end{equation}
where 
\begin{equation}
(\sigma_0,\sigma_1, \sigma_2,\sigma_3,\sigma_4)=\bigg(\frac{\lambda_0}{2m_4},\frac{\lambda_1}{m_3},\frac{2\lambda_2}{m_2},0,\frac{\lambda_0}{2 m_0}\bigg) \ .
\end{equation}
The form of the variation \eqref{eq:deltaSon} is the same as for the action of forms with ordinary kinetic terms. Since the three-form variation vanish, we should add a total derivative term giving a non-zero variation. A candidate term appears if one integrates by parts the action
\begin{equation}\label{eq:SB1}
    S_{B1} =\alpha_1 \int \dd\left[\epsilon_{AB}C_1^A\wedge C_3^B +\frac{\lambda_t}{4\kappa^2\lambda_1} (C_1^+-C_1^-)\wedge \Omega_3\right] \ .
\end{equation}
This action is not invariant off-shell under the zero-form symmetry \eqref{eq:gaugtr}, however, one can confirm that it is invariant on-shell. The variation of the on-shell action with the new terms for the one-form is still of the form \eqref{eq:deltaSon} with $\sigma_1'=\frac{\lambda_1+\alpha_1}{m_3}$. The variation for the three-form is 
\begin{equation}
    \delta S_\textrm{on-shell} =  \alpha_1 \int \dd\left[ C_1^+\wedge  \delta C_3^-- C_1^-\wedge  \delta C_3^+\right]\ .
\end{equation}
Another candidate is to add the total derivative term
\begin{equation}
    S_{B2} =-\frac{\sigma_3'}{2} \int \dd\left[\delta_{AB} C_3^A\wedge \star \dd C_3^B\right].
\end{equation}
In this case the variation is the same as for the other forms \eqref{eq:deltaSon}. The right choice for the coefficients turns out to be
\begin{equation}
    \sigma_3'=-\frac{4}{9}\frac{L^2}{2\kappa^2 g_w^2}\ , \qquad \alpha_1=-\frac{12}{g_w^2}\lambda_1\ .
\end{equation}

We relegate the details of the holographic renormalization in the Supplemental Material. The counterterm action takes the expected form of a sum of mass and kinetic terms of the fields evaluated at the cutoff, except for a term $\alpha_1 \int_{\partial AdS_5} \epsilon_{AB} c_1^A\wedge (C_3^B-\frac{3}{2}\Omega_3)$. This term depends on the source $c_\mu^\pm$ rather than on the pullback of the one-form $C_\mu^\pm$ and makes the whole action gauge-invariant off-shell.

%%%%%%%%%%%%%%%%%%%%%%%%%%%%%%%%%%%%%%%%%%%
\subsection{Expectation values}
%%%%%%%%%%%%%%%%%%%%%%%%%%%%%%%%%%%%%%%%%%%

The expectation value of the bilinear operators is determined by the variation $\corr{\delta I_p}=\delta S_\textrm{on-shell}+\delta S_{B1}+\delta S_{B2}+\delta S_{ct}$. We obtain\footnote{The contact terms can be modified by finite counterterms, see Supplemental Material.}
\begin{equation}
\!\!\!\!\! \corr{\cJ_p^\pm}=\frac{\sigma_p'}{L}\left(2 j_p^\pm-\frac{L^2}{2}\Box \hat c_p^\pm\right)+\delta_{p,3}\corr{\cJ_3^\pm}_{\alpha_1},
\end{equation}
where $\hat c_p^\pm=c_p^\pm$ for $p$ even, $\hat c_1^\pm=c_1^{\perp\pm}$ and $\hat c_3^\pm=0$. The additional term in the three-form is
\begin{equation}
    \corr{\cJ_3^\pm}_{\alpha_1}=\mp \alpha_1 \star_4 \left( j_1^{\perp\mp}- \frac{L^2}{4} \Box c_1^{\perp\mp}\right) \ .
\end{equation}
On the other hand, the spin current obtained from the variation with respect to the spin connection is
\begin{equation}
    \corr{\cS_3}=-\frac{2L}{\kappa^2 g_\omega^2}S_3-\frac{3}{2}\left(\corr{\cJ_3^+}_{\alpha_1}+\corr{\cJ_3^-}_{\alpha_1} \right) \ .
\end{equation}
Using \eqref{eq:j3coef}, one can check that the conditions $\corr{\cJ_3}=\frac{2}{3}\corr{\cS_3}$ and $\corr{\cJ_3^{\pm\parallel}}=\pm \star_4 \corr{\cJ_1^\mp}$ are satisfied.

%%%%%%%%%%%%%%%%%%%%%%%%%%%%%%%%%%%%%%%%%%%
\section{Nieh-Yan term in holography}
%%%%%%%%%%%%%%%%%%%%%%%%%%%%%%%%%%%%%%%%%%%

Our analysis reveals that the Hodge duality relations \eqref{eq:vevduality} for the odd forms are not satisfied in general. Defining $\corr{\cK_1^\pm}=\pm\star_4 \corr{\cJ_3^\mp}$, the corresponding axial and vector combinations are
\begin{equation}
    \corr{\cK_1}=\corr{\cJ_1}\ , \ \corr{\cK_{1,5}}=\corr{\cJ_{1,5}}-\frac{2i}{3}\star_4 \corr{\cS_3} \ . 
\end{equation}
Thus, while the Hodge dual vector current is conserved, the axial current acquires a longitudinal component
\begin{equation}
    \dd\star_4 \corr{ \cK_1}=0\ , \ \dd\star_4 \corr{ \cK_{1\,5}}=\frac{2i}{3}\dd\corr{\cS_3} \ .
\end{equation}
This introduces a temperature-dependent NY anomaly term. At non-zero temperature the dual geometry can be taken to be the $AdS_5$ black brane \eqref{eq:AdS5met}.
The regular solution for a zero-form of mass $M_0^2L^2=-3$ in this geometry is, at lowest order in momentum
\begin{equation}
     \Omega_0=\omega_0^{(0)}\frac{z}{L} \frac{4 \sqrt{\pi }}{\Gamma \left(\frac{1}{4}\right)^2} K\left(\frac{1}{2}-\frac{z^2}{2 z_h^2}\right) \ ,
\end{equation}
with $K(X)$ the complete elliptic integral of the first kind. This leads to
\begin{equation}
    S_3=\frac{3 g_w^2\kappa^2}{4L^3}c_\textrm{NY} \omega_3^{(0)} \ ,\quad c_\textrm{NY}=\frac{4 \sqrt{2} \pi ^3  \Gamma \left(\frac{3}{4}\right) }{9\Gamma \left(\frac{1}{4}\right)^3}\frac{L^3}{\kappa^2} T^2 \ .
\end{equation}

\begin{acknowledgments}
\emph{Acknowledgements.}---%
We thank Karl Landsteiner and Teemu Ojanen for useful discussions. C.~H. is partially supported by the Agencia Estatal de Investigaci\'on and the Ministerio de Ciencia, Innovaci\'on y Universidades
through the Spanish grant PID2021-123021NB-I00. N.~J. has been supported in part by Research Council of Finland grant no. 354533. 
\end{acknowledgments}

\bibliography{refs}

\begin{thebibliography}{73}%
\makeatletter
\providecommand \@ifxundefined [1]{%
 \@ifx{#1\undefined}
}%
\providecommand \@ifnum [1]{%
 \ifnum #1\expandafter \@firstoftwo
 \else \expandafter \@secondoftwo
 \fi
}%
\providecommand \@ifx [1]{%
 \ifx #1\expandafter \@firstoftwo
 \else \expandafter \@secondoftwo
 \fi
}%
\providecommand \natexlab [1]{#1}%
\providecommand \enquote  [1]{``#1''}%
\providecommand \bibnamefont  [1]{#1}%
\providecommand \bibfnamefont [1]{#1}%
\providecommand \citenamefont [1]{#1}%
\providecommand \href@noop [0]{\@secondoftwo}%
\providecommand \href [0]{\begingroup \@sanitize@url \@href}%
\providecommand \@href[1]{\@@startlink{#1}\@@href}%
\providecommand \@@href[1]{\endgroup#1\@@endlink}%
\providecommand \@sanitize@url [0]{\catcode `\\12\catcode `\$12\catcode
  `\&12\catcode `\#12\catcode `\^12\catcode `\_12\catcode `\%12\relax}%
\providecommand \@@startlink[1]{}%
\providecommand \@@endlink[0]{}%
\providecommand \url  [0]{\begingroup\@sanitize@url \@url }%
\providecommand \@url [1]{\endgroup\@href {#1}{\urlprefix }}%
\providecommand \urlprefix  [0]{URL }%
\providecommand \Eprint [0]{\href }%
\providecommand \doibase [0]{https://doi.org/}%
\providecommand \selectlanguage [0]{\@gobble}%
\providecommand \bibinfo  [0]{\@secondoftwo}%
\providecommand \bibfield  [0]{\@secondoftwo}%
\providecommand \translation [1]{[#1]}%
\providecommand \BibitemOpen [0]{}%
\providecommand \bibitemStop [0]{}%
\providecommand \bibitemNoStop [0]{.\EOS\space}%
\providecommand \EOS [0]{\spacefactor3000\relax}%
\providecommand \BibitemShut  [1]{\csname bibitem#1\endcsname}%
\let\auto@bib@innerbib\@empty
%</preamble>
\bibitem [{\citenamefont {Yan}\ and\ \citenamefont {Felser}(2017)}]{Yan_2017}%
  \BibitemOpen
  \bibfield  {author} {\bibinfo {author} {\bibfnamefont {B.}~\bibnamefont
  {Yan}}\ and\ \bibinfo {author} {\bibfnamefont {C.}~\bibnamefont {Felser}},\
  }\bibfield  {title} {\bibinfo {title} {Topological materials: Weyl
  semimetals},\ }\href
  {https://doi.org/10.1146/annurev-conmatphys-031016-025458} {\bibfield
  {journal} {\bibinfo  {journal} {Annual Review of Condensed Matter Physics}\
  }\textbf {\bibinfo {volume} {8}},\ \bibinfo {pages} {337–354} (\bibinfo
  {year} {2017})}\BibitemShut {NoStop}%
\bibitem [{\citenamefont {Volovik}(2006)}]{Volovik:2003fe}%
  \BibitemOpen
  \bibfield  {author} {\bibinfo {author} {\bibfnamefont {G.~E.}\ \bibnamefont
  {Volovik}},\ }\href@noop {} {\emph {\bibinfo {title} {{The Universe in a
  helium droplet}}}},\ Vol.\ \bibinfo {volume} {117}\ (\bibinfo {year}
  {2006})\BibitemShut {NoStop}%
\bibitem [{\citenamefont {Meng}\ and\ \citenamefont
  {Balents}(2012)}]{Meng_2012}%
  \BibitemOpen
  \bibfield  {author} {\bibinfo {author} {\bibfnamefont {T.}~\bibnamefont
  {Meng}}\ and\ \bibinfo {author} {\bibfnamefont {L.}~\bibnamefont {Balents}},\
  }\bibfield  {title} {\bibinfo {title} {Weyl superconductors},\ }\bibfield
  {journal} {\bibinfo  {journal} {Physical Review B}\ }\textbf {\bibinfo
  {volume} {86}},\ \href {https://doi.org/10.1103/physrevb.86.054504}
  {10.1103/physrevb.86.054504} (\bibinfo {year} {2012})\BibitemShut {NoStop}%
\bibitem [{\citenamefont {Chernodub}\ \emph {et~al.}(2022)\citenamefont
  {Chernodub}, \citenamefont {Ferreiros}, \citenamefont {Grushin},
  \citenamefont {Landsteiner},\ and\ \citenamefont
  {Vozmediano}}]{Chernodub:2021nff}%
  \BibitemOpen
  \bibfield  {author} {\bibinfo {author} {\bibfnamefont {M.~N.}\ \bibnamefont
  {Chernodub}}, \bibinfo {author} {\bibfnamefont {Y.}~\bibnamefont
  {Ferreiros}}, \bibinfo {author} {\bibfnamefont {A.~G.}\ \bibnamefont
  {Grushin}}, \bibinfo {author} {\bibfnamefont {K.}~\bibnamefont
  {Landsteiner}},\ and\ \bibinfo {author} {\bibfnamefont {M.~A.~H.}\
  \bibnamefont {Vozmediano}},\ }\bibfield  {title} {\bibinfo {title} {{Thermal
  transport, geometry, and anomalies}},\ }\href
  {https://doi.org/10.1016/j.physrep.2022.06.002} {\bibfield  {journal}
  {\bibinfo  {journal} {Phys. Rept.}\ }\textbf {\bibinfo {volume} {977}},\
  \bibinfo {pages} {1} (\bibinfo {year} {2022})},\ \Eprint
  {https://arxiv.org/abs/2110.05471} {arXiv:2110.05471 [cond-mat.mes-hall]}
  \BibitemShut {NoStop}%
\bibitem [{\citenamefont {Fukushima}\ \emph {et~al.}(2008)\citenamefont
  {Fukushima}, \citenamefont {Kharzeev},\ and\ \citenamefont
  {Warringa}}]{Fukushima:2008xe}%
  \BibitemOpen
  \bibfield  {author} {\bibinfo {author} {\bibfnamefont {K.}~\bibnamefont
  {Fukushima}}, \bibinfo {author} {\bibfnamefont {D.~E.}\ \bibnamefont
  {Kharzeev}},\ and\ \bibinfo {author} {\bibfnamefont {H.~J.}\ \bibnamefont
  {Warringa}},\ }\bibfield  {title} {\bibinfo {title} {{The Chiral Magnetic
  Effect}},\ }\href {https://doi.org/10.1103/PhysRevD.78.074033} {\bibfield
  {journal} {\bibinfo  {journal} {Phys. Rev. D}\ }\textbf {\bibinfo {volume}
  {78}},\ \bibinfo {pages} {074033} (\bibinfo {year} {2008})},\ \Eprint
  {https://arxiv.org/abs/0808.3382} {arXiv:0808.3382 [hep-ph]} \BibitemShut
  {NoStop}%
\bibitem [{\citenamefont {Fukushima}\ \emph {et~al.}(2010)\citenamefont
  {Fukushima}, \citenamefont {Kharzeev},\ and\ \citenamefont
  {Warringa}}]{Fukushima:2010vw}%
  \BibitemOpen
  \bibfield  {author} {\bibinfo {author} {\bibfnamefont {K.}~\bibnamefont
  {Fukushima}}, \bibinfo {author} {\bibfnamefont {D.~E.}\ \bibnamefont
  {Kharzeev}},\ and\ \bibinfo {author} {\bibfnamefont {H.~J.}\ \bibnamefont
  {Warringa}},\ }\bibfield  {title} {\bibinfo {title} {{Real-time dynamics of
  the Chiral Magnetic Effect}},\ }\href
  {https://doi.org/10.1103/PhysRevLett.104.212001} {\bibfield  {journal}
  {\bibinfo  {journal} {Phys. Rev. Lett.}\ }\textbf {\bibinfo {volume} {104}},\
  \bibinfo {pages} {212001} (\bibinfo {year} {2010})},\ \Eprint
  {https://arxiv.org/abs/1002.2495} {arXiv:1002.2495 [hep-ph]} \BibitemShut
  {NoStop}%
\bibitem [{\citenamefont {Kharzeev}(2014)}]{Kharzeev:2013ffa}%
  \BibitemOpen
  \bibfield  {author} {\bibinfo {author} {\bibfnamefont {D.~E.}\ \bibnamefont
  {Kharzeev}},\ }\bibfield  {title} {\bibinfo {title} {{The Chiral Magnetic
  Effect and Anomaly-Induced Transport}},\ }\href
  {https://doi.org/10.1016/j.ppnp.2014.01.002} {\bibfield  {journal} {\bibinfo
  {journal} {Prog. Part. Nucl. Phys.}\ }\textbf {\bibinfo {volume} {75}},\
  \bibinfo {pages} {133} (\bibinfo {year} {2014})},\ \Eprint
  {https://arxiv.org/abs/1312.3348} {arXiv:1312.3348 [hep-ph]} \BibitemShut
  {NoStop}%
\bibitem [{\citenamefont {Abelev}\ \emph {et~al.}(2009)\citenamefont {Abelev}
  \emph {et~al.}}]{STAR:2009wot}%
  \BibitemOpen
  \bibfield  {author} {\bibinfo {author} {\bibfnamefont {B.~I.}\ \bibnamefont
  {Abelev}} \emph {et~al.} (\bibinfo {collaboration} {STAR}),\ }\bibfield
  {title} {\bibinfo {title} {{Azimuthal Charged-Particle Correlations and
  Possible Local Strong Parity Violation}},\ }\href
  {https://doi.org/10.1103/PhysRevLett.103.251601} {\bibfield  {journal}
  {\bibinfo  {journal} {Phys. Rev. Lett.}\ }\textbf {\bibinfo {volume} {103}},\
  \bibinfo {pages} {251601} (\bibinfo {year} {2009})},\ \Eprint
  {https://arxiv.org/abs/0909.1739} {arXiv:0909.1739 [nucl-ex]} \BibitemShut
  {NoStop}%
\bibitem [{\citenamefont {Li}\ \emph {et~al.}(2016{\natexlab{a}})\citenamefont
  {Li}, \citenamefont {Kharzeev}, \citenamefont {Zhang}, \citenamefont {Huang},
  \citenamefont {Pletikosic}, \citenamefont {Fedorov}, \citenamefont {Zhong},
  \citenamefont {Schneeloch}, \citenamefont {Gu},\ and\ \citenamefont
  {Valla}}]{Li:2014bha}%
  \BibitemOpen
  \bibfield  {author} {\bibinfo {author} {\bibfnamefont {Q.}~\bibnamefont
  {Li}}, \bibinfo {author} {\bibfnamefont {D.~E.}\ \bibnamefont {Kharzeev}},
  \bibinfo {author} {\bibfnamefont {C.}~\bibnamefont {Zhang}}, \bibinfo
  {author} {\bibfnamefont {Y.}~\bibnamefont {Huang}}, \bibinfo {author}
  {\bibfnamefont {I.}~\bibnamefont {Pletikosic}}, \bibinfo {author}
  {\bibfnamefont {A.~V.}\ \bibnamefont {Fedorov}}, \bibinfo {author}
  {\bibfnamefont {R.~D.}\ \bibnamefont {Zhong}}, \bibinfo {author}
  {\bibfnamefont {J.~A.}\ \bibnamefont {Schneeloch}}, \bibinfo {author}
  {\bibfnamefont {G.~D.}\ \bibnamefont {Gu}},\ and\ \bibinfo {author}
  {\bibfnamefont {T.}~\bibnamefont {Valla}},\ }\bibfield  {title} {\bibinfo
  {title} {{Observation of the chiral magnetic effect in ZrTe5}},\ }\href
  {https://doi.org/10.1038/nphys3648} {\bibfield  {journal} {\bibinfo
  {journal} {Nature Phys.}\ }\textbf {\bibinfo {volume} {12}},\ \bibinfo
  {pages} {550} (\bibinfo {year} {2016}{\natexlab{a}})},\ \Eprint
  {https://arxiv.org/abs/1412.6543} {arXiv:1412.6543 [cond-mat.str-el]}
  \BibitemShut {NoStop}%
\bibitem [{\citenamefont {Kharzeev}\ \emph {et~al.}(2024)\citenamefont
  {Kharzeev}, \citenamefont {Liao},\ and\ \citenamefont
  {Tribedy}}]{Kharzeev:2024zzm}%
  \BibitemOpen
  \bibfield  {author} {\bibinfo {author} {\bibfnamefont {D.~E.}\ \bibnamefont
  {Kharzeev}}, \bibinfo {author} {\bibfnamefont {J.}~\bibnamefont {Liao}},\
  and\ \bibinfo {author} {\bibfnamefont {P.}~\bibnamefont {Tribedy}},\
  }\bibfield  {title} {\bibinfo {title} {{Chiral Magnetic Effect in Heavy Ion
  Collisions: The Present and Future}},\ }\href@noop {} {\  (\bibinfo {year}
  {2024})},\ \Eprint {https://arxiv.org/abs/2405.05427} {arXiv:2405.05427
  [nucl-th]} \BibitemShut {NoStop}%
\bibitem [{\citenamefont {Nielsen}\ and\ \citenamefont
  {Ninomiya}(1983)}]{Nielsen:1983rb}%
  \BibitemOpen
  \bibfield  {author} {\bibinfo {author} {\bibfnamefont {H.~B.}\ \bibnamefont
  {Nielsen}}\ and\ \bibinfo {author} {\bibfnamefont {M.}~\bibnamefont
  {Ninomiya}},\ }\bibfield  {title} {\bibinfo {title} {{ADLER-BELL-JACKIW
  ANOMALY AND WEYL FERMIONS IN CRYSTAL}},\ }\href
  {https://doi.org/10.1016/0370-2693(83)91529-0} {\bibfield  {journal}
  {\bibinfo  {journal} {Phys. Lett. B}\ }\textbf {\bibinfo {volume} {130}},\
  \bibinfo {pages} {389} (\bibinfo {year} {1983})}\BibitemShut {NoStop}%
\bibitem [{\citenamefont {Son}\ and\ \citenamefont {Spivak}(2013)}]{Son_2013}%
  \BibitemOpen
  \bibfield  {author} {\bibinfo {author} {\bibfnamefont {D.~T.}\ \bibnamefont
  {Son}}\ and\ \bibinfo {author} {\bibfnamefont {B.~Z.}\ \bibnamefont
  {Spivak}},\ }\bibfield  {title} {\bibinfo {title} {Chiral anomaly and
  classical negative magnetoresistance of weyl metals},\ }\bibfield  {journal}
  {\bibinfo  {journal} {Physical Review B}\ }\textbf {\bibinfo {volume} {88}},\
  \href {https://doi.org/10.1103/physrevb.88.104412}
  {10.1103/physrevb.88.104412} (\bibinfo {year} {2013})\BibitemShut {NoStop}%
\bibitem [{\citenamefont {Lundgren}\ \emph {et~al.}(2014)\citenamefont
  {Lundgren}, \citenamefont {Laurell},\ and\ \citenamefont
  {Fiete}}]{Lundgren:2014hra}%
  \BibitemOpen
  \bibfield  {author} {\bibinfo {author} {\bibfnamefont {R.}~\bibnamefont
  {Lundgren}}, \bibinfo {author} {\bibfnamefont {P.}~\bibnamefont {Laurell}},\
  and\ \bibinfo {author} {\bibfnamefont {G.~A.}\ \bibnamefont {Fiete}},\
  }\bibfield  {title} {\bibinfo {title} {{Thermoelectric properties of Weyl and
  Dirac semimetals}},\ }\href {https://doi.org/10.1103/PhysRevB.90.165115}
  {\bibfield  {journal} {\bibinfo  {journal} {Phys. Rev. B}\ }\textbf {\bibinfo
  {volume} {90}},\ \bibinfo {pages} {165115} (\bibinfo {year} {2014})},\
  \Eprint {https://arxiv.org/abs/1407.1435} {arXiv:1407.1435 [cond-mat.str-el]}
  \BibitemShut {NoStop}%
\bibitem [{\citenamefont {Burkov}(2015)}]{Burkov_2015}%
  \BibitemOpen
  \bibfield  {author} {\bibinfo {author} {\bibfnamefont {A.~A.}\ \bibnamefont
  {Burkov}},\ }\bibfield  {title} {\bibinfo {title} {Negative longitudinal
  magnetoresistance in dirac and weyl metals},\ }\bibfield  {journal} {\bibinfo
   {journal} {Physical Review B}\ }\textbf {\bibinfo {volume} {91}},\ \href
  {https://doi.org/10.1103/physrevb.91.245157} {10.1103/physrevb.91.245157}
  (\bibinfo {year} {2015})\BibitemShut {NoStop}%
\bibitem [{\citenamefont {Spivak}\ and\ \citenamefont
  {Andreev}(2016)}]{Spivak_2016}%
  \BibitemOpen
  \bibfield  {author} {\bibinfo {author} {\bibfnamefont {B.~Z.}\ \bibnamefont
  {Spivak}}\ and\ \bibinfo {author} {\bibfnamefont {A.~V.}\ \bibnamefont
  {Andreev}},\ }\bibfield  {title} {\bibinfo {title} {Magnetotransport
  phenomena related to the chiral anomaly in weyl semimetals},\ }\bibfield
  {journal} {\bibinfo  {journal} {Physical Review B}\ }\textbf {\bibinfo
  {volume} {93}},\ \href {https://doi.org/10.1103/physrevb.93.085107}
  {10.1103/physrevb.93.085107} (\bibinfo {year} {2016})\BibitemShut {NoStop}%
\bibitem [{\citenamefont {Lucas}\ \emph {et~al.}(2016)\citenamefont {Lucas},
  \citenamefont {Davison},\ and\ \citenamefont {Sachdev}}]{Lucas:2016omy}%
  \BibitemOpen
  \bibfield  {author} {\bibinfo {author} {\bibfnamefont {A.}~\bibnamefont
  {Lucas}}, \bibinfo {author} {\bibfnamefont {R.~A.}\ \bibnamefont {Davison}},\
  and\ \bibinfo {author} {\bibfnamefont {S.}~\bibnamefont {Sachdev}},\
  }\bibfield  {title} {\bibinfo {title} {{Hydrodynamic theory of thermoelectric
  transport and negative magnetoresistance in Weyl semimetals}},\ }\href
  {https://doi.org/10.1073/pnas.1608881113} {\bibfield  {journal} {\bibinfo
  {journal} {Proc. Nat. Acad. Sci.}\ }\textbf {\bibinfo {volume} {113}},\
  \bibinfo {pages} {9463} (\bibinfo {year} {2016})},\ \Eprint
  {https://arxiv.org/abs/1604.08598} {arXiv:1604.08598 [cond-mat.str-el]}
  \BibitemShut {NoStop}%
\bibitem [{\citenamefont {Kim}\ \emph {et~al.}(2013)\citenamefont {Kim},
  \citenamefont {Kim}, \citenamefont {Wang}, \citenamefont {Sasaki},
  \citenamefont {Satoh}, \citenamefont {Ohnishi}, \citenamefont {Kitaura},
  \citenamefont {Yang},\ and\ \citenamefont {Li}}]{Kim_2013}%
  \BibitemOpen
  \bibfield  {author} {\bibinfo {author} {\bibfnamefont {H.-J.}\ \bibnamefont
  {Kim}}, \bibinfo {author} {\bibfnamefont {K.-S.}\ \bibnamefont {Kim}},
  \bibinfo {author} {\bibfnamefont {J.-F.}\ \bibnamefont {Wang}}, \bibinfo
  {author} {\bibfnamefont {M.}~\bibnamefont {Sasaki}}, \bibinfo {author}
  {\bibfnamefont {N.}~\bibnamefont {Satoh}}, \bibinfo {author} {\bibfnamefont
  {A.}~\bibnamefont {Ohnishi}}, \bibinfo {author} {\bibfnamefont
  {M.}~\bibnamefont {Kitaura}}, \bibinfo {author} {\bibfnamefont
  {M.}~\bibnamefont {Yang}},\ and\ \bibinfo {author} {\bibfnamefont
  {L.}~\bibnamefont {Li}},\ }\bibfield  {title} {\bibinfo {title} {Dirac versus
  weyl fermions in topological insulators: Adler-bell-jackiw anomaly in
  transport phenomena},\ }\bibfield  {journal} {\bibinfo  {journal} {Physical
  Review Letters}\ }\textbf {\bibinfo {volume} {111}},\ \href
  {https://doi.org/10.1103/physrevlett.111.246603}
  {10.1103/physrevlett.111.246603} (\bibinfo {year} {2013})\BibitemShut
  {NoStop}%
\bibitem [{\citenamefont {Xiong}\ \emph {et~al.}(2015)\citenamefont {Xiong},
  \citenamefont {Kushwaha}, \citenamefont {Liang}, \citenamefont {Krizan},
  \citenamefont {Hirschberger}, \citenamefont {Wang}, \citenamefont {Cava},\
  and\ \citenamefont {Ong}}]{doi:10.1126/science.aac6089}%
  \BibitemOpen
  \bibfield  {author} {\bibinfo {author} {\bibfnamefont {J.}~\bibnamefont
  {Xiong}}, \bibinfo {author} {\bibfnamefont {S.~K.}\ \bibnamefont {Kushwaha}},
  \bibinfo {author} {\bibfnamefont {T.}~\bibnamefont {Liang}}, \bibinfo
  {author} {\bibfnamefont {J.~W.}\ \bibnamefont {Krizan}}, \bibinfo {author}
  {\bibfnamefont {M.}~\bibnamefont {Hirschberger}}, \bibinfo {author}
  {\bibfnamefont {W.}~\bibnamefont {Wang}}, \bibinfo {author} {\bibfnamefont
  {R.~J.}\ \bibnamefont {Cava}},\ and\ \bibinfo {author} {\bibfnamefont
  {N.~P.}\ \bibnamefont {Ong}},\ }\bibfield  {title} {\bibinfo {title}
  {Evidence for the chiral anomaly in the dirac semimetal na<sub>3</sub>bi},\
  }\href {https://doi.org/10.1126/science.aac6089} {\bibfield  {journal}
  {\bibinfo  {journal} {Science}\ }\textbf {\bibinfo {volume} {350}},\ \bibinfo
  {pages} {413} (\bibinfo {year} {2015})},\ \Eprint
  {https://arxiv.org/abs/https://www.science.org/doi/pdf/10.1126/science.aac6089}
  {https://www.science.org/doi/pdf/10.1126/science.aac6089} \BibitemShut
  {NoStop}%
\bibitem [{\citenamefont {Huang}\ \emph {et~al.}(2015)\citenamefont {Huang},
  \citenamefont {Zhao}, \citenamefont {Long}, \citenamefont {Wang},
  \citenamefont {Chen}, \citenamefont {Yang}, \citenamefont {Liang},
  \citenamefont {Xue}, \citenamefont {Weng}, \citenamefont {Fang},
  \citenamefont {Dai},\ and\ \citenamefont {Chen}}]{Huang_2015}%
  \BibitemOpen
  \bibfield  {author} {\bibinfo {author} {\bibfnamefont {X.}~\bibnamefont
  {Huang}}, \bibinfo {author} {\bibfnamefont {L.}~\bibnamefont {Zhao}},
  \bibinfo {author} {\bibfnamefont {Y.}~\bibnamefont {Long}}, \bibinfo {author}
  {\bibfnamefont {P.}~\bibnamefont {Wang}}, \bibinfo {author} {\bibfnamefont
  {D.}~\bibnamefont {Chen}}, \bibinfo {author} {\bibfnamefont {Z.}~\bibnamefont
  {Yang}}, \bibinfo {author} {\bibfnamefont {H.}~\bibnamefont {Liang}},
  \bibinfo {author} {\bibfnamefont {M.}~\bibnamefont {Xue}}, \bibinfo {author}
  {\bibfnamefont {H.}~\bibnamefont {Weng}}, \bibinfo {author} {\bibfnamefont
  {Z.}~\bibnamefont {Fang}}, \bibinfo {author} {\bibfnamefont {X.}~\bibnamefont
  {Dai}},\ and\ \bibinfo {author} {\bibfnamefont {G.}~\bibnamefont {Chen}},\
  }\bibfield  {title} {\bibinfo {title} {Observation of the
  chiral-anomaly-induced negative magnetoresistance in 3d weyl semimetal
  taas},\ }\bibfield  {journal} {\bibinfo  {journal} {Physical Review X}\
  }\textbf {\bibinfo {volume} {5}},\ \href
  {https://doi.org/10.1103/physrevx.5.031023} {10.1103/physrevx.5.031023}
  (\bibinfo {year} {2015})\BibitemShut {NoStop}%
\bibitem [{\citenamefont {Li}\ \emph {et~al.}(2015)\citenamefont {Li},
  \citenamefont {Wang}, \citenamefont {Liu}, \citenamefont {Wang},
  \citenamefont {Liao},\ and\ \citenamefont {Yu}}]{Li_2015}%
  \BibitemOpen
  \bibfield  {author} {\bibinfo {author} {\bibfnamefont {C.-Z.}\ \bibnamefont
  {Li}}, \bibinfo {author} {\bibfnamefont {L.-X.}\ \bibnamefont {Wang}},
  \bibinfo {author} {\bibfnamefont {H.}~\bibnamefont {Liu}}, \bibinfo {author}
  {\bibfnamefont {J.}~\bibnamefont {Wang}}, \bibinfo {author} {\bibfnamefont
  {Z.-M.}\ \bibnamefont {Liao}},\ and\ \bibinfo {author} {\bibfnamefont
  {D.-P.}\ \bibnamefont {Yu}},\ }\bibfield  {title} {\bibinfo {title} {Giant
  negative magnetoresistance induced by the chiral anomaly in individual cd3as2
  nanowires},\ }\bibfield  {journal} {\bibinfo  {journal} {Nature
  Communications}\ }\textbf {\bibinfo {volume} {6}},\ \href
  {https://doi.org/10.1038/ncomms10137} {10.1038/ncomms10137} (\bibinfo {year}
  {2015})\BibitemShut {NoStop}%
\bibitem [{\citenamefont {Li}\ \emph {et~al.}(2016{\natexlab{b}})\citenamefont
  {Li}, \citenamefont {He}, \citenamefont {Lu}, \citenamefont {Zhang},
  \citenamefont {Liu}, \citenamefont {Ma}, \citenamefont {Fan}, \citenamefont
  {Shen},\ and\ \citenamefont {Wang}}]{Li_2016}%
  \BibitemOpen
  \bibfield  {author} {\bibinfo {author} {\bibfnamefont {H.}~\bibnamefont
  {Li}}, \bibinfo {author} {\bibfnamefont {H.}~\bibnamefont {He}}, \bibinfo
  {author} {\bibfnamefont {H.-Z.}\ \bibnamefont {Lu}}, \bibinfo {author}
  {\bibfnamefont {H.}~\bibnamefont {Zhang}}, \bibinfo {author} {\bibfnamefont
  {H.}~\bibnamefont {Liu}}, \bibinfo {author} {\bibfnamefont {R.}~\bibnamefont
  {Ma}}, \bibinfo {author} {\bibfnamefont {Z.}~\bibnamefont {Fan}}, \bibinfo
  {author} {\bibfnamefont {S.-Q.}\ \bibnamefont {Shen}},\ and\ \bibinfo
  {author} {\bibfnamefont {J.}~\bibnamefont {Wang}},\ }\bibfield  {title}
  {\bibinfo {title} {Negative magnetoresistance in dirac semimetal cd3as2},\
  }\bibfield  {journal} {\bibinfo  {journal} {Nature Communications}\ }\textbf
  {\bibinfo {volume} {7}},\ \href {https://doi.org/10.1038/ncomms10301}
  {10.1038/ncomms10301} (\bibinfo {year} {2016}{\natexlab{b}})\BibitemShut
  {NoStop}%
\bibitem [{\citenamefont {Zhang}\ \emph {et~al.}(2016)\citenamefont {Zhang},
  \citenamefont {Xu}, \citenamefont {Belopolski}, \citenamefont {Yuan},
  \citenamefont {Lin}, \citenamefont {Tong}, \citenamefont {Bian},
  \citenamefont {Alidoust}, \citenamefont {Lee}, \citenamefont {Huang},
  \citenamefont {Chang}, \citenamefont {Chang}, \citenamefont {Hsu},
  \citenamefont {Jeng}, \citenamefont {Neupane}, \citenamefont {Sanchez},
  \citenamefont {Zheng}, \citenamefont {Wang}, \citenamefont {Lin},
  \citenamefont {Zhang}, \citenamefont {Lu}, \citenamefont {Shen},
  \citenamefont {Neupert}, \citenamefont {Zahid~Hasan},\ and\ \citenamefont
  {Jia}}]{Zhang_2016}%
  \BibitemOpen
  \bibfield  {author} {\bibinfo {author} {\bibfnamefont {C.-L.}\ \bibnamefont
  {Zhang}}, \bibinfo {author} {\bibfnamefont {S.-Y.}\ \bibnamefont {Xu}},
  \bibinfo {author} {\bibfnamefont {I.}~\bibnamefont {Belopolski}}, \bibinfo
  {author} {\bibfnamefont {Z.}~\bibnamefont {Yuan}}, \bibinfo {author}
  {\bibfnamefont {Z.}~\bibnamefont {Lin}}, \bibinfo {author} {\bibfnamefont
  {B.}~\bibnamefont {Tong}}, \bibinfo {author} {\bibfnamefont {G.}~\bibnamefont
  {Bian}}, \bibinfo {author} {\bibfnamefont {N.}~\bibnamefont {Alidoust}},
  \bibinfo {author} {\bibfnamefont {C.-C.}\ \bibnamefont {Lee}}, \bibinfo
  {author} {\bibfnamefont {S.-M.}\ \bibnamefont {Huang}}, \bibinfo {author}
  {\bibfnamefont {T.-R.}\ \bibnamefont {Chang}}, \bibinfo {author}
  {\bibfnamefont {G.}~\bibnamefont {Chang}}, \bibinfo {author} {\bibfnamefont
  {C.-H.}\ \bibnamefont {Hsu}}, \bibinfo {author} {\bibfnamefont {H.-T.}\
  \bibnamefont {Jeng}}, \bibinfo {author} {\bibfnamefont {M.}~\bibnamefont
  {Neupane}}, \bibinfo {author} {\bibfnamefont {D.~S.}\ \bibnamefont
  {Sanchez}}, \bibinfo {author} {\bibfnamefont {H.}~\bibnamefont {Zheng}},
  \bibinfo {author} {\bibfnamefont {J.}~\bibnamefont {Wang}}, \bibinfo {author}
  {\bibfnamefont {H.}~\bibnamefont {Lin}}, \bibinfo {author} {\bibfnamefont
  {C.}~\bibnamefont {Zhang}}, \bibinfo {author} {\bibfnamefont {H.-Z.}\
  \bibnamefont {Lu}}, \bibinfo {author} {\bibfnamefont {S.-Q.}\ \bibnamefont
  {Shen}}, \bibinfo {author} {\bibfnamefont {T.}~\bibnamefont {Neupert}},
  \bibinfo {author} {\bibfnamefont {M.}~\bibnamefont {Zahid~Hasan}},\ and\
  \bibinfo {author} {\bibfnamefont {S.}~\bibnamefont {Jia}},\ }\bibfield
  {title} {\bibinfo {title} {Signatures of the adler–bell–jackiw chiral
  anomaly in a weyl fermion semimetal},\ }\bibfield  {journal} {\bibinfo
  {journal} {Nature Communications}\ }\textbf {\bibinfo {volume} {7}},\ \href
  {https://doi.org/10.1038/ncomms10735} {10.1038/ncomms10735} (\bibinfo {year}
  {2016})\BibitemShut {NoStop}%
\bibitem [{\citenamefont {Hirschberger}\ \emph {et~al.}(2016)\citenamefont
  {Hirschberger}, \citenamefont {Kushwaha}, \citenamefont {Wang}, \citenamefont
  {Gibson}, \citenamefont {Liang}, \citenamefont {Belvin}, \citenamefont
  {Bernevig}, \citenamefont {Cava},\ and\ \citenamefont
  {Ong}}]{Hirschberger_2016}%
  \BibitemOpen
  \bibfield  {author} {\bibinfo {author} {\bibfnamefont {M.}~\bibnamefont
  {Hirschberger}}, \bibinfo {author} {\bibfnamefont {S.}~\bibnamefont
  {Kushwaha}}, \bibinfo {author} {\bibfnamefont {Z.}~\bibnamefont {Wang}},
  \bibinfo {author} {\bibfnamefont {Q.}~\bibnamefont {Gibson}}, \bibinfo
  {author} {\bibfnamefont {S.}~\bibnamefont {Liang}}, \bibinfo {author}
  {\bibfnamefont {C.}~\bibnamefont {Belvin}}, \bibinfo {author} {\bibfnamefont
  {B.}~\bibnamefont {Bernevig}}, \bibinfo {author} {\bibfnamefont
  {R.}~\bibnamefont {Cava}},\ and\ \bibinfo {author} {\bibfnamefont
  {N.}~\bibnamefont {Ong}},\ }\bibfield  {title} {\bibinfo {title} {The chiral
  anomaly and thermopower of weyl fermions in the half-heusler gdptbi},\
  }\href {https://doi.org/10.1038/nmat4684} {\bibfield  {journal} {\bibinfo
  {journal} {Nature Materials}\ }\textbf {\bibinfo {volume} {15}},\ \bibinfo
  {pages} {1161–1165} (\bibinfo {year} {2016})}\BibitemShut {NoStop}%
\bibitem [{\citenamefont {Gooth}\ \emph {et~al.}(2017)\citenamefont {Gooth}
  \emph {et~al.}}]{Gooth:2017mbd}%
  \BibitemOpen
  \bibfield  {author} {\bibinfo {author} {\bibfnamefont {J.}~\bibnamefont
  {Gooth}} \emph {et~al.},\ }\bibfield  {title} {\bibinfo {title}
  {{Experimental signatures of the mixed axial-gravitational anomaly in the
  Weyl semimetal NbP}},\ }\href {https://doi.org/10.1038/nature23005}
  {\bibfield  {journal} {\bibinfo  {journal} {Nature}\ }\textbf {\bibinfo
  {volume} {547}},\ \bibinfo {pages} {324} (\bibinfo {year} {2017})},\ \Eprint
  {https://arxiv.org/abs/1703.10682} {arXiv:1703.10682 [cond-mat.mtrl-sci]}
  \BibitemShut {NoStop}%
\bibitem [{\citenamefont {de~Juan}\ \emph {et~al.}(2010)\citenamefont
  {de~Juan}, \citenamefont {Cortijo},\ and\ \citenamefont
  {Vozmediano}}]{deJuan:2009ldt}%
  \BibitemOpen
  \bibfield  {author} {\bibinfo {author} {\bibfnamefont {F.}~\bibnamefont
  {de~Juan}}, \bibinfo {author} {\bibfnamefont {A.}~\bibnamefont {Cortijo}},\
  and\ \bibinfo {author} {\bibfnamefont {M.~A.~H.}\ \bibnamefont
  {Vozmediano}},\ }\bibfield  {title} {\bibinfo {title} {{Dislocations and
  torsion in graphene and related systems}},\ }\href
  {https://doi.org/10.1016/j.nuclphysb.2009.11.012} {\bibfield  {journal}
  {\bibinfo  {journal} {Nucl. Phys. B}\ }\textbf {\bibinfo {volume} {828}},\
  \bibinfo {pages} {625} (\bibinfo {year} {2010})},\ \Eprint
  {https://arxiv.org/abs/0909.4068} {arXiv:0909.4068 [cond-mat.mes-hall]}
  \BibitemShut {NoStop}%
\bibitem [{\citenamefont {Hughes}\ \emph {et~al.}(2013)\citenamefont {Hughes},
  \citenamefont {Leigh},\ and\ \citenamefont {Parrikar}}]{Hughes:2012vg}%
  \BibitemOpen
  \bibfield  {author} {\bibinfo {author} {\bibfnamefont {T.~L.}\ \bibnamefont
  {Hughes}}, \bibinfo {author} {\bibfnamefont {R.~G.}\ \bibnamefont {Leigh}},\
  and\ \bibinfo {author} {\bibfnamefont {O.}~\bibnamefont {Parrikar}},\
  }\bibfield  {title} {\bibinfo {title} {{Torsional Anomalies, Hall Viscosity,
  and Bulk-boundary Correspondence in Topological States}},\ }\href
  {https://doi.org/10.1103/PhysRevD.88.025040} {\bibfield  {journal} {\bibinfo
  {journal} {Phys. Rev. D}\ }\textbf {\bibinfo {volume} {88}},\ \bibinfo
  {pages} {025040} (\bibinfo {year} {2013})},\ \Eprint
  {https://arxiv.org/abs/1211.6442} {arXiv:1211.6442 [hep-th]} \BibitemShut
  {NoStop}%
\bibitem [{\citenamefont {Parrikar}\ \emph {et~al.}(2014)\citenamefont
  {Parrikar}, \citenamefont {Hughes},\ and\ \citenamefont
  {Leigh}}]{Parrikar:2014usa}%
  \BibitemOpen
  \bibfield  {author} {\bibinfo {author} {\bibfnamefont {O.}~\bibnamefont
  {Parrikar}}, \bibinfo {author} {\bibfnamefont {T.~L.}\ \bibnamefont
  {Hughes}},\ and\ \bibinfo {author} {\bibfnamefont {R.~G.}\ \bibnamefont
  {Leigh}},\ }\bibfield  {title} {\bibinfo {title} {{Torsion, Parity-odd
  Response and Anomalies in Topological States}},\ }\href
  {https://doi.org/10.1103/PhysRevD.90.105004} {\bibfield  {journal} {\bibinfo
  {journal} {Phys. Rev. D}\ }\textbf {\bibinfo {volume} {90}},\ \bibinfo
  {pages} {105004} (\bibinfo {year} {2014})},\ \Eprint
  {https://arxiv.org/abs/1407.7043} {arXiv:1407.7043 [cond-mat.mes-hall]}
  \BibitemShut {NoStop}%
\bibitem [{\citenamefont {Ishihara}\ \emph {et~al.}(2019)\citenamefont
  {Ishihara}, \citenamefont {Mizushima}, \citenamefont {Tsuruta},\ and\
  \citenamefont {Fujimoto}}]{Ishihara_2019}%
  \BibitemOpen
  \bibfield  {author} {\bibinfo {author} {\bibfnamefont {Y.}~\bibnamefont
  {Ishihara}}, \bibinfo {author} {\bibfnamefont {T.}~\bibnamefont {Mizushima}},
  \bibinfo {author} {\bibfnamefont {A.}~\bibnamefont {Tsuruta}},\ and\ \bibinfo
  {author} {\bibfnamefont {S.}~\bibnamefont {Fujimoto}},\ }\bibfield  {title}
  {\bibinfo {title} {Torsional chiral magnetic effect due to skyrmion textures
  in a weyl superfluid},\ }\bibfield  {journal} {\bibinfo  {journal} {Physical
  Review B}\ }\textbf {\bibinfo {volume} {99}},\ \href
  {https://doi.org/10.1103/physrevb.99.024513} {10.1103/physrevb.99.024513}
  (\bibinfo {year} {2019})\BibitemShut {NoStop}%
\bibitem [{\citenamefont {Nieh}\ and\ \citenamefont {Yan}(1982)}]{Nieh:1981ww}%
  \BibitemOpen
  \bibfield  {author} {\bibinfo {author} {\bibfnamefont {H.~T.}\ \bibnamefont
  {Nieh}}\ and\ \bibinfo {author} {\bibfnamefont {M.~L.}\ \bibnamefont {Yan}},\
  }\bibfield  {title} {\bibinfo {title} {{An Identity in Riemann-cartan
  Geometry}},\ }\href {https://doi.org/10.1063/1.525379} {\bibfield  {journal}
  {\bibinfo  {journal} {J. Math. Phys.}\ }\textbf {\bibinfo {volume} {23}},\
  \bibinfo {pages} {373} (\bibinfo {year} {1982})}\BibitemShut {NoStop}%
\bibitem [{\citenamefont {Chandia}\ and\ \citenamefont
  {Zanelli}(1997)}]{Chandia:1997hu}%
  \BibitemOpen
  \bibfield  {author} {\bibinfo {author} {\bibfnamefont {O.}~\bibnamefont
  {Chandia}}\ and\ \bibinfo {author} {\bibfnamefont {J.}~\bibnamefont
  {Zanelli}},\ }\bibfield  {title} {\bibinfo {title} {{Topological invariants,
  instantons and chiral anomaly on spaces with torsion}},\ }\href
  {https://doi.org/10.1103/PhysRevD.55.7580} {\bibfield  {journal} {\bibinfo
  {journal} {Phys. Rev. D}\ }\textbf {\bibinfo {volume} {55}},\ \bibinfo
  {pages} {7580} (\bibinfo {year} {1997})},\ \Eprint
  {https://arxiv.org/abs/hep-th/9702025} {arXiv:hep-th/9702025} \BibitemShut
  {NoStop}%
\bibitem [{\citenamefont {Obukhov}\ \emph {et~al.}(1997)\citenamefont
  {Obukhov}, \citenamefont {Mielke}, \citenamefont {Budczies},\ and\
  \citenamefont {Hehl}}]{Obukhov:1997pz}%
  \BibitemOpen
  \bibfield  {author} {\bibinfo {author} {\bibfnamefont {Y.~N.}\ \bibnamefont
  {Obukhov}}, \bibinfo {author} {\bibfnamefont {E.~W.}\ \bibnamefont {Mielke}},
  \bibinfo {author} {\bibfnamefont {J.}~\bibnamefont {Budczies}},\ and\
  \bibinfo {author} {\bibfnamefont {F.~W.}\ \bibnamefont {Hehl}},\ }\bibfield
  {title} {\bibinfo {title} {{On the chiral anomaly in nonRiemannian
  space-times}},\ }\href {https://doi.org/10.1007/BF02551525} {\bibfield
  {journal} {\bibinfo  {journal} {Found. Phys.}\ }\textbf {\bibinfo {volume}
  {27}},\ \bibinfo {pages} {1221} (\bibinfo {year} {1997})},\ \Eprint
  {https://arxiv.org/abs/gr-qc/9702011} {arXiv:gr-qc/9702011} \BibitemShut
  {NoStop}%
\bibitem [{\citenamefont {Kreimer}\ and\ \citenamefont
  {Mielke}(2001)}]{Kreimer:1999yp}%
  \BibitemOpen
  \bibfield  {author} {\bibinfo {author} {\bibfnamefont {D.}~\bibnamefont
  {Kreimer}}\ and\ \bibinfo {author} {\bibfnamefont {E.~W.}\ \bibnamefont
  {Mielke}},\ }\bibfield  {title} {\bibinfo {title} {{Comment on: Topological
  invariants, instantons, and the chiral anomaly on spaces with torsion}},\
  }\href {https://doi.org/10.1103/PhysRevD.63.048501} {\bibfield  {journal}
  {\bibinfo  {journal} {Phys. Rev. D}\ }\textbf {\bibinfo {volume} {63}},\
  \bibinfo {pages} {048501} (\bibinfo {year} {2001})},\ \Eprint
  {https://arxiv.org/abs/gr-qc/9904071} {arXiv:gr-qc/9904071} \BibitemShut
  {NoStop}%
\bibitem [{\citenamefont {Chandia}\ and\ \citenamefont
  {Zanelli}(2001)}]{Chandia:1999az}%
  \BibitemOpen
  \bibfield  {author} {\bibinfo {author} {\bibfnamefont {O.}~\bibnamefont
  {Chandia}}\ and\ \bibinfo {author} {\bibfnamefont {J.}~\bibnamefont
  {Zanelli}},\ }\bibfield  {title} {\bibinfo {title} {{Reply to the comment by
  D. Kreimer and E. Mielke}},\ }\href
  {https://doi.org/10.1103/PhysRevD.63.048502} {\bibfield  {journal} {\bibinfo
  {journal} {Phys. Rev. D}\ }\textbf {\bibinfo {volume} {63}},\ \bibinfo
  {pages} {048502} (\bibinfo {year} {2001})},\ \Eprint
  {https://arxiv.org/abs/hep-th/9906165} {arXiv:hep-th/9906165} \BibitemShut
  {NoStop}%
\bibitem [{\citenamefont {Erdmenger}\ \emph
  {et~al.}(2024{\natexlab{a}})\citenamefont {Erdmenger}, \citenamefont
  {Matthaiakakis}, \citenamefont {Meyer},\ and\ \citenamefont
  {Vassilevich}}]{Erdmenger:2024zty}%
  \BibitemOpen
  \bibfield  {author} {\bibinfo {author} {\bibfnamefont {J.}~\bibnamefont
  {Erdmenger}}, \bibinfo {author} {\bibfnamefont {I.}~\bibnamefont
  {Matthaiakakis}}, \bibinfo {author} {\bibfnamefont {R.}~\bibnamefont
  {Meyer}},\ and\ \bibinfo {author} {\bibfnamefont {D.}~\bibnamefont
  {Vassilevich}},\ }\bibfield  {title} {\bibinfo {title} {{The chiral torsional
  anomaly and the Nieh-Yan invariant with and without boundaries}},\
  }\href@noop {} {\  (\bibinfo {year} {2024}{\natexlab{a}})},\ \Eprint
  {https://arxiv.org/abs/2409.06766} {arXiv:2409.06766 [hep-th]} \BibitemShut
  {NoStop}%
\bibitem [{\citenamefont {Sumiyoshi}\ and\ \citenamefont
  {Fujimoto}(2016)}]{Sumiyoshi:2015eda}%
  \BibitemOpen
  \bibfield  {author} {\bibinfo {author} {\bibfnamefont {H.}~\bibnamefont
  {Sumiyoshi}}\ and\ \bibinfo {author} {\bibfnamefont {S.}~\bibnamefont
  {Fujimoto}},\ }\bibfield  {title} {\bibinfo {title} {{Torsional Chiral
  Magnetic Effect in a Weyl Semimetal with a Topological Defect}},\ }\href
  {https://doi.org/10.1103/PhysRevLett.116.166601} {\bibfield  {journal}
  {\bibinfo  {journal} {Phys. Rev. Lett.}\ }\textbf {\bibinfo {volume} {116}},\
  \bibinfo {pages} {166601} (\bibinfo {year} {2016})},\ \Eprint
  {https://arxiv.org/abs/1509.03981} {arXiv:1509.03981 [cond-mat.mes-hall]}
  \BibitemShut {NoStop}%
\bibitem [{\citenamefont {Ferreiros}\ \emph {et~al.}(2019)\citenamefont
  {Ferreiros}, \citenamefont {Kedem}, \citenamefont {Bergholtz},\ and\
  \citenamefont {Bardarson}}]{Ferreiros:2018udw}%
  \BibitemOpen
  \bibfield  {author} {\bibinfo {author} {\bibfnamefont {Y.}~\bibnamefont
  {Ferreiros}}, \bibinfo {author} {\bibfnamefont {Y.}~\bibnamefont {Kedem}},
  \bibinfo {author} {\bibfnamefont {E.~J.}\ \bibnamefont {Bergholtz}},\ and\
  \bibinfo {author} {\bibfnamefont {J.~H.}\ \bibnamefont {Bardarson}},\
  }\bibfield  {title} {\bibinfo {title} {{Mixed axial-torsional anomaly in Weyl
  semimetals}},\ }\href {https://doi.org/10.1103/PhysRevLett.122.056601}
  {\bibfield  {journal} {\bibinfo  {journal} {Phys. Rev. Lett.}\ }\textbf
  {\bibinfo {volume} {122}},\ \bibinfo {pages} {056601} (\bibinfo {year}
  {2019})},\ \Eprint {https://arxiv.org/abs/1808.08241} {arXiv:1808.08241
  [cond-mat.mes-hall]} \BibitemShut {NoStop}%
\bibitem [{\citenamefont {Huang}\ \emph {et~al.}(2019)\citenamefont {Huang},
  \citenamefont {Li}, \citenamefont {Zhou},\ and\ \citenamefont
  {Zhang}}]{Huang_2019}%
  \BibitemOpen
  \bibfield  {author} {\bibinfo {author} {\bibfnamefont {Z.-M.}\ \bibnamefont
  {Huang}}, \bibinfo {author} {\bibfnamefont {L.}~\bibnamefont {Li}}, \bibinfo
  {author} {\bibfnamefont {J.}~\bibnamefont {Zhou}},\ and\ \bibinfo {author}
  {\bibfnamefont {H.-H.}\ \bibnamefont {Zhang}},\ }\bibfield  {title} {\bibinfo
  {title} {Torsional response and liouville anomaly in weyl semimetals with
  dislocations},\ }\bibfield  {journal} {\bibinfo  {journal} {Physical Review
  B}\ }\textbf {\bibinfo {volume} {99}},\ \href
  {https://doi.org/10.1103/physrevb.99.155152} {10.1103/physrevb.99.155152}
  (\bibinfo {year} {2019})\BibitemShut {NoStop}%
\bibitem [{\citenamefont {Huang}\ and\ \citenamefont
  {Han}(2020)}]{Huang:2020ypv}%
  \BibitemOpen
  \bibfield  {author} {\bibinfo {author} {\bibfnamefont {Z.-M.}\ \bibnamefont
  {Huang}}\ and\ \bibinfo {author} {\bibfnamefont {B.}~\bibnamefont {Han}},\
  }\bibfield  {title} {\bibinfo {title} {{Torsional Anomalies and
  Bulk-Dislocation Correspondence in Weyl Systems}},\ }\href@noop {} {\
  (\bibinfo {year} {2020})},\ \Eprint {https://arxiv.org/abs/2003.04853}
  {arXiv:2003.04853 [cond-mat.mes-hall]} \BibitemShut {NoStop}%
\bibitem [{\citenamefont {Huang}\ \emph
  {et~al.}(2020{\natexlab{a}})\citenamefont {Huang}, \citenamefont {Han},\ and\
  \citenamefont {Stone}}]{Huang_2020b}%
  \BibitemOpen
  \bibfield  {author} {\bibinfo {author} {\bibfnamefont {Z.-M.}\ \bibnamefont
  {Huang}}, \bibinfo {author} {\bibfnamefont {B.}~\bibnamefont {Han}},\ and\
  \bibinfo {author} {\bibfnamefont {M.}~\bibnamefont {Stone}},\ }\bibfield
  {title} {\bibinfo {title} {Hamiltonian approach to the torsional anomalies
  and its dimensional ladder},\ }\bibfield  {journal} {\bibinfo  {journal}
  {Physical Review B}\ }\textbf {\bibinfo {volume} {101}},\ \href
  {https://doi.org/10.1103/physrevb.101.165201} {10.1103/physrevb.101.165201}
  (\bibinfo {year} {2020}{\natexlab{a}})\BibitemShut {NoStop}%
\bibitem [{\citenamefont {Nissinen}(2020)}]{Nissinen:2019kld}%
  \BibitemOpen
  \bibfield  {author} {\bibinfo {author} {\bibfnamefont {J.}~\bibnamefont
  {Nissinen}},\ }\bibfield  {title} {\bibinfo {title} {{Emergent spacetime and
  gravitational Nieh-Yan anomaly in chiral $p+ip$ Weyl superfluids and
  superconductors}},\ }\href {https://doi.org/10.1103/PhysRevLett.124.117002}
  {\bibfield  {journal} {\bibinfo  {journal} {Phys. Rev. Lett.}\ }\textbf
  {\bibinfo {volume} {124}},\ \bibinfo {pages} {117002} (\bibinfo {year}
  {2020})},\ \Eprint {https://arxiv.org/abs/1909.05846} {arXiv:1909.05846
  [cond-mat.supr-con]} \BibitemShut {NoStop}%
\bibitem [{\citenamefont {Laurila}\ and\ \citenamefont
  {Nissinen}(2020)}]{Laurila:2020yll}%
  \BibitemOpen
  \bibfield  {author} {\bibinfo {author} {\bibfnamefont {S.}~\bibnamefont
  {Laurila}}\ and\ \bibinfo {author} {\bibfnamefont {J.}~\bibnamefont
  {Nissinen}},\ }\bibfield  {title} {\bibinfo {title} {{Torsional Landau levels
  and geometric anomalies in condensed matter Weyl systems}},\ }\href
  {https://doi.org/10.1103/PhysRevB.102.235163} {\bibfield  {journal} {\bibinfo
   {journal} {Phys. Rev. B}\ }\textbf {\bibinfo {volume} {102}},\ \bibinfo
  {pages} {235163} (\bibinfo {year} {2020})},\ \Eprint
  {https://arxiv.org/abs/2007.10682} {arXiv:2007.10682 [cond-mat.str-el]}
  \BibitemShut {NoStop}%
\bibitem [{\citenamefont {Nissinen}\ and\ \citenamefont
  {Volovik}(2019)}]{Nissinen:2019wmh}%
  \BibitemOpen
  \bibfield  {author} {\bibinfo {author} {\bibfnamefont {J.}~\bibnamefont
  {Nissinen}}\ and\ \bibinfo {author} {\bibfnamefont {G.~E.}\ \bibnamefont
  {Volovik}},\ }\bibfield  {title} {\bibinfo {title} {{On thermal
  Nieh\textendash{}Yan anomaly in topological Weyl materials}},\ }\href
  {https://doi.org/10.1134/S0021364019240020} {\bibfield  {journal} {\bibinfo
  {journal} {Pisma Zh. Eksp. Teor. Fiz.}\ }\textbf {\bibinfo {volume} {110}},\
  \bibinfo {pages} {797} (\bibinfo {year} {2019})},\ \Eprint
  {https://arxiv.org/abs/1911.03382} {arXiv:1911.03382 [cond-mat.str-el]}
  \BibitemShut {NoStop}%
\bibitem [{\citenamefont {Nissinen}\ and\ \citenamefont
  {Volovik}(2020)}]{Nissinen:2019mkw}%
  \BibitemOpen
  \bibfield  {author} {\bibinfo {author} {\bibfnamefont {J.}~\bibnamefont
  {Nissinen}}\ and\ \bibinfo {author} {\bibfnamefont {G.~E.}\ \bibnamefont
  {Volovik}},\ }\bibfield  {title} {\bibinfo {title} {{Thermal Nieh-Yan anomaly
  in Weyl superfluids}},\ }\href
  {https://doi.org/10.1103/PhysRevResearch.2.033269} {\bibfield  {journal}
  {\bibinfo  {journal} {Phys. Rev. Res.}\ }\textbf {\bibinfo {volume} {2}},\
  \bibinfo {pages} {033269} (\bibinfo {year} {2020})},\ \Eprint
  {https://arxiv.org/abs/1909.08936} {arXiv:1909.08936 [cond-mat.str-el]}
  \BibitemShut {NoStop}%
\bibitem [{\citenamefont {Huang}\ \emph
  {et~al.}(2020{\natexlab{b}})\citenamefont {Huang}, \citenamefont {Han},\ and\
  \citenamefont {Stone}}]{Huang_2020}%
  \BibitemOpen
  \bibfield  {author} {\bibinfo {author} {\bibfnamefont {Z.-M.}\ \bibnamefont
  {Huang}}, \bibinfo {author} {\bibfnamefont {B.}~\bibnamefont {Han}},\ and\
  \bibinfo {author} {\bibfnamefont {M.}~\bibnamefont {Stone}},\ }\bibfield
  {title} {\bibinfo {title} {Nieh-yan anomaly: Torsional landau levels, central
  charge, and anomalous thermal hall effect},\ }\bibfield  {journal} {\bibinfo
  {journal} {Physical Review B}\ }\textbf {\bibinfo {volume} {101}},\ \href
  {https://doi.org/10.1103/physrevb.101.125201} {10.1103/physrevb.101.125201}
  (\bibinfo {year} {2020}{\natexlab{b}})\BibitemShut {NoStop}%
\bibitem [{\citenamefont {Liang}\ and\ \citenamefont
  {Ojanen}(2020)}]{Liang_2020}%
  \BibitemOpen
  \bibfield  {author} {\bibinfo {author} {\bibfnamefont {L.}~\bibnamefont
  {Liang}}\ and\ \bibinfo {author} {\bibfnamefont {T.}~\bibnamefont {Ojanen}},\
  }\bibfield  {title} {\bibinfo {title} {Topological magnetotorsional effect in
  weyl semimetals},\ }\bibfield  {journal} {\bibinfo  {journal} {Physical
  Review Research}\ }\textbf {\bibinfo {volume} {2}},\ \href
  {https://doi.org/10.1103/physrevresearch.2.022016}
  {10.1103/physrevresearch.2.022016} (\bibinfo {year} {2020})\BibitemShut
  {NoStop}%
\bibitem [{\citenamefont {Khaidukov}\ and\ \citenamefont
  {Zubkov}(2018)}]{Khaidukov:2018oat}%
  \BibitemOpen
  \bibfield  {author} {\bibinfo {author} {\bibfnamefont {Z.~V.}\ \bibnamefont
  {Khaidukov}}\ and\ \bibinfo {author} {\bibfnamefont {M.~A.}\ \bibnamefont
  {Zubkov}},\ }\bibfield  {title} {\bibinfo {title} {{Chiral torsional
  effect}},\ }\href {https://doi.org/10.1134/S0021364018220046} {\bibfield
  {journal} {\bibinfo  {journal} {JETP Lett.}\ }\textbf {\bibinfo {volume}
  {108}},\ \bibinfo {pages} {670} (\bibinfo {year} {2018})},\ \Eprint
  {https://arxiv.org/abs/1812.00970} {arXiv:1812.00970 [cond-mat.mes-hall]}
  \BibitemShut {NoStop}%
\bibitem [{\citenamefont {Imaki}\ and\ \citenamefont
  {Yamamoto}(2019)}]{Imaki:2019ite}%
  \BibitemOpen
  \bibfield  {author} {\bibinfo {author} {\bibfnamefont {S.}~\bibnamefont
  {Imaki}}\ and\ \bibinfo {author} {\bibfnamefont {A.}~\bibnamefont
  {Yamamoto}},\ }\bibfield  {title} {\bibinfo {title} {{Lattice field theory
  with torsion}},\ }\href {https://doi.org/10.1103/PhysRevD.100.054509}
  {\bibfield  {journal} {\bibinfo  {journal} {Phys. Rev. D}\ }\textbf {\bibinfo
  {volume} {100}},\ \bibinfo {pages} {054509} (\bibinfo {year} {2019})},\
  \Eprint {https://arxiv.org/abs/1906.02406} {arXiv:1906.02406 [hep-lat]}
  \BibitemShut {NoStop}%
\bibitem [{\citenamefont {Imaki}\ and\ \citenamefont
  {Qiu}(2020)}]{Imaki:2020csc}%
  \BibitemOpen
  \bibfield  {author} {\bibinfo {author} {\bibfnamefont {S.}~\bibnamefont
  {Imaki}}\ and\ \bibinfo {author} {\bibfnamefont {Z.}~\bibnamefont {Qiu}},\
  }\bibfield  {title} {\bibinfo {title} {{Chiral torsional effect with finite
  temperature, density and curvature}},\ }\href
  {https://doi.org/10.1103/PhysRevD.102.016001} {\bibfield  {journal} {\bibinfo
   {journal} {Phys. Rev. D}\ }\textbf {\bibinfo {volume} {102}},\ \bibinfo
  {pages} {016001} (\bibinfo {year} {2020})},\ \Eprint
  {https://arxiv.org/abs/2004.11899} {arXiv:2004.11899 [hep-th]} \BibitemShut
  {NoStop}%
\bibitem [{\citenamefont {Nissinen}\ and\ \citenamefont
  {Volovik}(2022)}]{Nissinen:2021gke}%
  \BibitemOpen
  \bibfield  {author} {\bibinfo {author} {\bibfnamefont {J.}~\bibnamefont
  {Nissinen}}\ and\ \bibinfo {author} {\bibfnamefont {G.~E.}\ \bibnamefont
  {Volovik}},\ }\bibfield  {title} {\bibinfo {title} {{Anomalous chiral
  transport with vorticity and torsion: Cancellation of two mixed gravitational
  anomaly currents in rotating chiral p+ip Weyl condensates}},\ }\href
  {https://doi.org/10.1103/PhysRevD.106.045022} {\bibfield  {journal} {\bibinfo
   {journal} {Phys. Rev. D}\ }\textbf {\bibinfo {volume} {106}},\ \bibinfo
  {pages} {045022} (\bibinfo {year} {2022})},\ \Eprint
  {https://arxiv.org/abs/2111.08639} {arXiv:2111.08639 [cond-mat.supr-con]}
  \BibitemShut {NoStop}%
\bibitem [{\citenamefont {Ferreiros}\ and\ \citenamefont
  {Landsteiner}(2021)}]{Ferreiros:2020uda}%
  \BibitemOpen
  \bibfield  {author} {\bibinfo {author} {\bibfnamefont {Y.}~\bibnamefont
  {Ferreiros}}\ and\ \bibinfo {author} {\bibfnamefont {K.}~\bibnamefont
  {Landsteiner}},\ }\bibfield  {title} {\bibinfo {title} {{On chiral responses
  to geometric torsion}},\ }\href
  {https://doi.org/10.1016/j.physletb.2021.136419} {\bibfield  {journal}
  {\bibinfo  {journal} {Phys. Lett. B}\ }\textbf {\bibinfo {volume} {819}},\
  \bibinfo {pages} {136419} (\bibinfo {year} {2021})},\ \Eprint
  {https://arxiv.org/abs/2011.10535} {arXiv:2011.10535 [cond-mat.mes-hall]}
  \BibitemShut {NoStop}%
\bibitem [{\citenamefont {Amitani}\ and\ \citenamefont
  {Nishida}(2023)}]{Amitani:2022xev}%
  \BibitemOpen
  \bibfield  {author} {\bibinfo {author} {\bibfnamefont {T.}~\bibnamefont
  {Amitani}}\ and\ \bibinfo {author} {\bibfnamefont {Y.}~\bibnamefont
  {Nishida}},\ }\bibfield  {title} {\bibinfo {title} {{Torsion-induced chiral
  magnetic current in equilibrium}},\ }\href
  {https://doi.org/10.1016/j.aop.2022.169181} {\bibfield  {journal} {\bibinfo
  {journal} {Annals Phys.}\ }\textbf {\bibinfo {volume} {448}},\ \bibinfo
  {pages} {169181} (\bibinfo {year} {2023})},\ \Eprint
  {https://arxiv.org/abs/2204.13415} {arXiv:2204.13415 [hep-th]} \BibitemShut
  {NoStop}%
\bibitem [{\citenamefont {Valle}\ and\ \citenamefont
  {Vazquez-Mozo}(2022)}]{Valle:2021nfv}%
  \BibitemOpen
  \bibfield  {author} {\bibinfo {author} {\bibfnamefont {M.}~\bibnamefont
  {Valle}}\ and\ \bibinfo {author} {\bibfnamefont {M.~A.}\ \bibnamefont
  {Vazquez-Mozo}},\ }\bibfield  {title} {\bibinfo {title} {{On Nieh-Yan
  transport}},\ }\href {https://doi.org/10.1007/JHEP03(2022)177} {\bibfield
  {journal} {\bibinfo  {journal} {JHEP}\ }\textbf {\bibinfo {volume} {03}},\
  \bibinfo {pages} {177}},\ \Eprint {https://arxiv.org/abs/2112.02003}
  {arXiv:2112.02003 [hep-th]} \BibitemShut {NoStop}%
\bibitem [{\citenamefont {Valle}\ and\ \citenamefont
  {Vazquez-Mozo}(2024)}]{Valle:2023cqo}%
  \BibitemOpen
  \bibfield  {author} {\bibinfo {author} {\bibfnamefont {M.}~\bibnamefont
  {Valle}}\ and\ \bibinfo {author} {\bibfnamefont {M.~A.}\ \bibnamefont
  {Vazquez-Mozo}},\ }\bibfield  {title} {\bibinfo {title} {{Torsional
  constitutive relations at finite temperature}},\ }\href
  {https://doi.org/10.1007/JHEP02(2024)185} {\bibfield  {journal} {\bibinfo
  {journal} {JHEP}\ }\textbf {\bibinfo {volume} {02}},\ \bibinfo {pages}
  {185}},\ \Eprint {https://arxiv.org/abs/2312.12081} {arXiv:2312.12081
  [hep-th]} \BibitemShut {NoStop}%
\bibitem [{\citenamefont {Landsteiner}\ \emph {et~al.}(2011)\citenamefont
  {Landsteiner}, \citenamefont {Megias}, \citenamefont {Melgar},\ and\
  \citenamefont {Pena-Benitez}}]{Landsteiner:2011iq}%
  \BibitemOpen
  \bibfield  {author} {\bibinfo {author} {\bibfnamefont {K.}~\bibnamefont
  {Landsteiner}}, \bibinfo {author} {\bibfnamefont {E.}~\bibnamefont {Megias}},
  \bibinfo {author} {\bibfnamefont {L.}~\bibnamefont {Melgar}},\ and\ \bibinfo
  {author} {\bibfnamefont {F.}~\bibnamefont {Pena-Benitez}},\ }\bibfield
  {title} {\bibinfo {title} {{Holographic Gravitational Anomaly and Chiral
  Vortical Effect}},\ }\href {https://doi.org/10.1007/JHEP09(2011)121}
  {\bibfield  {journal} {\bibinfo  {journal} {JHEP}\ }\textbf {\bibinfo
  {volume} {09}},\ \bibinfo {pages} {121}},\ \Eprint
  {https://arxiv.org/abs/1107.0368} {arXiv:1107.0368 [hep-th]} \BibitemShut
  {NoStop}%
\bibitem [{\citenamefont {Braguta}\ \emph {et~al.}(2013)\citenamefont
  {Braguta}, \citenamefont {Chernodub}, \citenamefont {Landsteiner},
  \citenamefont {Polikarpov},\ and\ \citenamefont
  {Ulybyshev}}]{Braguta:2013loa}%
  \BibitemOpen
  \bibfield  {author} {\bibinfo {author} {\bibfnamefont {V.}~\bibnamefont
  {Braguta}}, \bibinfo {author} {\bibfnamefont {M.~N.}\ \bibnamefont
  {Chernodub}}, \bibinfo {author} {\bibfnamefont {K.}~\bibnamefont
  {Landsteiner}}, \bibinfo {author} {\bibfnamefont {M.~I.}\ \bibnamefont
  {Polikarpov}},\ and\ \bibinfo {author} {\bibfnamefont {M.~V.}\ \bibnamefont
  {Ulybyshev}},\ }\bibfield  {title} {\bibinfo {title} {{Numerical evidence of
  the axial magnetic effect}},\ }\href
  {https://doi.org/10.1103/PhysRevD.88.071501} {\bibfield  {journal} {\bibinfo
  {journal} {Phys. Rev. D}\ }\textbf {\bibinfo {volume} {88}},\ \bibinfo
  {pages} {071501} (\bibinfo {year} {2013})},\ \Eprint
  {https://arxiv.org/abs/1303.6266} {arXiv:1303.6266 [hep-lat]} \BibitemShut
  {NoStop}%
\bibitem [{\citenamefont {Leigh}\ \emph {et~al.}(2009)\citenamefont {Leigh},
  \citenamefont {Hoang},\ and\ \citenamefont {Petkou}}]{Leigh:2008tt}%
  \BibitemOpen
  \bibfield  {author} {\bibinfo {author} {\bibfnamefont {R.~G.}\ \bibnamefont
  {Leigh}}, \bibinfo {author} {\bibfnamefont {N.~N.}\ \bibnamefont {Hoang}},\
  and\ \bibinfo {author} {\bibfnamefont {A.~C.}\ \bibnamefont {Petkou}},\
  }\bibfield  {title} {\bibinfo {title} {{Torsion and the Gravity Dual of
  Parity Symmetry Breaking in AdS(4) / CFT(3) Holography}},\ }\href
  {https://doi.org/10.1088/1126-6708/2009/03/033} {\bibfield  {journal}
  {\bibinfo  {journal} {JHEP}\ }\textbf {\bibinfo {volume} {03}},\ \bibinfo
  {pages} {033}},\ \Eprint {https://arxiv.org/abs/0809.5258} {arXiv:0809.5258
  [hep-th]} \BibitemShut {NoStop}%
\bibitem [{\citenamefont {Petkou}(2010)}]{Petkou:2010ve}%
  \BibitemOpen
  \bibfield  {author} {\bibinfo {author} {\bibfnamefont {A.~C.}\ \bibnamefont
  {Petkou}},\ }\bibfield  {title} {\bibinfo {title} {{Torsional degrees of
  freedom in AdS4/CFT3}}\ }(\bibinfo {year} {2010})\ \Eprint
  {https://arxiv.org/abs/1004.1640} {arXiv:1004.1640 [hep-th]} \BibitemShut
  {NoStop}%
\bibitem [{\citenamefont {Gallegos}\ and\ \citenamefont
  {G\"ursoy}(2020)}]{Gallegos:2020otk}%
  \BibitemOpen
  \bibfield  {author} {\bibinfo {author} {\bibfnamefont {A.~D.}\ \bibnamefont
  {Gallegos}}\ and\ \bibinfo {author} {\bibfnamefont {U.}~\bibnamefont
  {G\"ursoy}},\ }\bibfield  {title} {\bibinfo {title} {{Holographic spin
  liquids and Lovelock Chern-Simons gravity}},\ }\href
  {https://doi.org/10.1007/JHEP11(2020)151} {\bibfield  {journal} {\bibinfo
  {journal} {JHEP}\ }\textbf {\bibinfo {volume} {11}},\ \bibinfo {pages}
  {151}},\ \Eprint {https://arxiv.org/abs/2004.05148} {arXiv:2004.05148
  [hep-th]} \BibitemShut {NoStop}%
\bibitem [{\citenamefont {Gallegos}\ \emph {et~al.}(2023)\citenamefont
  {Gallegos}, \citenamefont {Gursoy},\ and\ \citenamefont
  {Yarom}}]{Gallegos:2022jow}%
  \BibitemOpen
  \bibfield  {author} {\bibinfo {author} {\bibfnamefont {A.~D.}\ \bibnamefont
  {Gallegos}}, \bibinfo {author} {\bibfnamefont {U.}~\bibnamefont {Gursoy}},\
  and\ \bibinfo {author} {\bibfnamefont {A.}~\bibnamefont {Yarom}},\ }\bibfield
   {title} {\bibinfo {title} {{Hydrodynamics, spin currents and torsion}},\
  }\href {https://doi.org/10.1007/JHEP05(2023)139} {\bibfield  {journal}
  {\bibinfo  {journal} {JHEP}\ }\textbf {\bibinfo {volume} {05}},\ \bibinfo
  {pages} {139}},\ \Eprint {https://arxiv.org/abs/2203.05044} {arXiv:2203.05044
  [hep-th]} \BibitemShut {NoStop}%
\bibitem [{\citenamefont {Erdmenger}\ \emph {et~al.}(2023)\citenamefont
  {Erdmenger}, \citenamefont {He\ss{}}, \citenamefont {Matthaiakakis},\ and\
  \citenamefont {Meyer}}]{Erdmenger:2022nhz}%
  \BibitemOpen
  \bibfield  {author} {\bibinfo {author} {\bibfnamefont {J.}~\bibnamefont
  {Erdmenger}}, \bibinfo {author} {\bibfnamefont {B.}~\bibnamefont {He\ss{}}},
  \bibinfo {author} {\bibfnamefont {I.}~\bibnamefont {Matthaiakakis}},\ and\
  \bibinfo {author} {\bibfnamefont {R.}~\bibnamefont {Meyer}},\ }\bibfield
  {title} {\bibinfo {title} {{Universal Gibbons-Hawking-York term for theories
  with curvature, torsion and non-metricity}},\ }\href
  {https://doi.org/10.21468/SciPostPhys.14.5.099} {\bibfield  {journal}
  {\bibinfo  {journal} {SciPost Phys.}\ }\textbf {\bibinfo {volume} {14}},\
  \bibinfo {pages} {099} (\bibinfo {year} {2023})},\ \Eprint
  {https://arxiv.org/abs/2211.02064} {arXiv:2211.02064 [hep-th]} \BibitemShut
  {NoStop}%
\bibitem [{\citenamefont {Erdmenger}\ \emph
  {et~al.}(2024{\natexlab{b}})\citenamefont {Erdmenger}, \citenamefont
  {He\ss{}}, \citenamefont {Meyer},\ and\ \citenamefont
  {Matthaiakakis}}]{Erdmenger:2023hne}%
  \BibitemOpen
  \bibfield  {author} {\bibinfo {author} {\bibfnamefont {J.}~\bibnamefont
  {Erdmenger}}, \bibinfo {author} {\bibfnamefont {B.}~\bibnamefont {He\ss{}}},
  \bibinfo {author} {\bibfnamefont {R.}~\bibnamefont {Meyer}},\ and\ \bibinfo
  {author} {\bibfnamefont {I.}~\bibnamefont {Matthaiakakis}},\ }\bibfield
  {title} {\bibinfo {title} {{Gibbons-Hawking-York boundary terms and the
  generalized geometrical trinity of gravity}},\ }\href
  {https://doi.org/10.1103/PhysRevD.110.066002} {\bibfield  {journal} {\bibinfo
   {journal} {Phys. Rev. D}\ }\textbf {\bibinfo {volume} {110}},\ \bibinfo
  {pages} {066002} (\bibinfo {year} {2024}{\natexlab{b}})},\ \Eprint
  {https://arxiv.org/abs/2304.06752} {arXiv:2304.06752 [hep-th]} \BibitemShut
  {NoStop}%
\bibitem [{\citenamefont {Cartwright}\ \emph {et~al.}(2024)\citenamefont
  {Cartwright}, \citenamefont {Gallegos}, \citenamefont {G\"ursoy},
  \citenamefont {Klein},\ and\ \citenamefont {Yarom}}]{Cartwright:2024dcj}%
  \BibitemOpen
  \bibfield  {author} {\bibinfo {author} {\bibfnamefont {C.}~\bibnamefont
  {Cartwright}}, \bibinfo {author} {\bibfnamefont {D.}~\bibnamefont
  {Gallegos}}, \bibinfo {author} {\bibfnamefont {U.}~\bibnamefont {G\"ursoy}},
  \bibinfo {author} {\bibfnamefont {R.}~\bibnamefont {Klein}},\ and\ \bibinfo
  {author} {\bibfnamefont {A.}~\bibnamefont {Yarom}},\ }\bibfield  {title}
  {\bibinfo {title} {{A supersymmetric spin current}},\ }\href@noop {} {\
  (\bibinfo {year} {2024})},\ \Eprint {https://arxiv.org/abs/2408.04399}
  {arXiv:2408.04399 [hep-th]} \BibitemShut {NoStop}%
\bibitem [{\citenamefont {Adamczyk}\ \emph {et~al.}(2017)\citenamefont
  {Adamczyk} \emph {et~al.}}]{STAR:2017ckg}%
  \BibitemOpen
  \bibfield  {author} {\bibinfo {author} {\bibfnamefont {L.}~\bibnamefont
  {Adamczyk}} \emph {et~al.} (\bibinfo {collaboration} {STAR}),\ }\bibfield
  {title} {\bibinfo {title} {{Global $\Lambda$ hyperon polarization in nuclear
  collisions: evidence for the most vortical fluid}},\ }\href
  {https://doi.org/10.1038/nature23004} {\bibfield  {journal} {\bibinfo
  {journal} {Nature}\ }\textbf {\bibinfo {volume} {548}},\ \bibinfo {pages}
  {62} (\bibinfo {year} {2017})},\ \Eprint {https://arxiv.org/abs/1701.06657}
  {arXiv:1701.06657 [nucl-ex]} \BibitemShut {NoStop}%
\bibitem [{\citenamefont {Adam}\ \emph {et~al.}(2019)\citenamefont {Adam} \emph
  {et~al.}}]{STAR:2019erd}%
  \BibitemOpen
  \bibfield  {author} {\bibinfo {author} {\bibfnamefont {J.}~\bibnamefont
  {Adam}} \emph {et~al.} (\bibinfo {collaboration} {STAR}),\ }\bibfield
  {title} {\bibinfo {title} {{Polarization of $\Lambda$ ($\bar{\Lambda}$)
  hyperons along the beam direction in Au+Au collisions at $\sqrt{s_{_{NN}}}$ =
  200 GeV}},\ }\href {https://doi.org/10.1103/PhysRevLett.123.132301}
  {\bibfield  {journal} {\bibinfo  {journal} {Phys. Rev. Lett.}\ }\textbf
  {\bibinfo {volume} {123}},\ \bibinfo {pages} {132301} (\bibinfo {year}
  {2019})},\ \Eprint {https://arxiv.org/abs/1905.11917} {arXiv:1905.11917
  [nucl-ex]} \BibitemShut {NoStop}%
\bibitem [{\citenamefont {Acharya}\ \emph {et~al.}(2020)\citenamefont {Acharya}
  \emph {et~al.}}]{ALICE:2019aid}%
  \BibitemOpen
  \bibfield  {author} {\bibinfo {author} {\bibfnamefont {S.}~\bibnamefont
  {Acharya}} \emph {et~al.} (\bibinfo {collaboration} {ALICE}),\ }\bibfield
  {title} {\bibinfo {title} {{Evidence of Spin-Orbital Angular Momentum
  Interactions in Relativistic Heavy-Ion Collisions}},\ }\href
  {https://doi.org/10.1103/PhysRevLett.125.012301} {\bibfield  {journal}
  {\bibinfo  {journal} {Phys. Rev. Lett.}\ }\textbf {\bibinfo {volume} {125}},\
  \bibinfo {pages} {012301} (\bibinfo {year} {2020})},\ \Eprint
  {https://arxiv.org/abs/1910.14408} {arXiv:1910.14408 [nucl-ex]} \BibitemShut
  {NoStop}%
\bibitem [{\citenamefont {Adam}\ \emph {et~al.}(2021)\citenamefont {Adam} \emph
  {et~al.}}]{STAR:2020xbm}%
  \BibitemOpen
  \bibfield  {author} {\bibinfo {author} {\bibfnamefont {J.}~\bibnamefont
  {Adam}} \emph {et~al.} (\bibinfo {collaboration} {STAR}),\ }\bibfield
  {title} {\bibinfo {title} {{Global Polarization of $\Xi$ and $\Omega$
  Hyperons in Au+Au Collisions at $\sqrt {s_{NN}}$ = 200 GeV}},\ }\href
  {https://doi.org/10.1103/PhysRevLett.126.162301} {\bibfield  {journal}
  {\bibinfo  {journal} {Phys. Rev. Lett.}\ }\textbf {\bibinfo {volume} {126}},\
  \bibinfo {pages} {162301} (\bibinfo {year} {2021})},\ \bibinfo {note}
  {[Erratum: Phys.Rev.Lett. 131, 089901 (2023)]},\ \Eprint
  {https://arxiv.org/abs/2012.13601} {arXiv:2012.13601 [nucl-ex]} \BibitemShut
  {NoStop}%
\bibitem [{\citenamefont {Ma\~nes}\ \emph {et~al.}(2021)\citenamefont
  {Ma\~nes}, \citenamefont {Valle},\ and\ \citenamefont
  {V\'azquez-Mozo}}]{Manes:2020zdd}%
  \BibitemOpen
  \bibfield  {author} {\bibinfo {author} {\bibfnamefont {J.~L.}\ \bibnamefont
  {Ma\~nes}}, \bibinfo {author} {\bibfnamefont {M.}~\bibnamefont {Valle}},\
  and\ \bibinfo {author} {\bibfnamefont {M.~A.}\ \bibnamefont
  {V\'azquez-Mozo}},\ }\bibfield  {title} {\bibinfo {title} {{Chiral torsional
  effects in anomalous fluids in thermal equilibrium}},\ }\href
  {https://doi.org/10.1007/JHEP05(2021)209} {\bibfield  {journal} {\bibinfo
  {journal} {JHEP}\ }\textbf {\bibinfo {volume} {05}},\ \bibinfo {pages}
  {209}},\ \Eprint {https://arxiv.org/abs/2012.08449} {arXiv:2012.08449
  [hep-th]} \BibitemShut {NoStop}%
\bibitem [{\citenamefont {Alvares}\ \emph {et~al.}(2011)\citenamefont
  {Alvares}, \citenamefont {Hoyos},\ and\ \citenamefont
  {Karch}}]{Alvares:2011wb}%
  \BibitemOpen
  \bibfield  {author} {\bibinfo {author} {\bibfnamefont {R.}~\bibnamefont
  {Alvares}}, \bibinfo {author} {\bibfnamefont {C.}~\bibnamefont {Hoyos}},\
  and\ \bibinfo {author} {\bibfnamefont {A.}~\bibnamefont {Karch}},\ }\bibfield
   {title} {\bibinfo {title} {{An improved model of vector mesons in
  holographic QCD}},\ }\href {https://doi.org/10.1103/PhysRevD.84.095020}
  {\bibfield  {journal} {\bibinfo  {journal} {Phys. Rev. D}\ }\textbf {\bibinfo
  {volume} {84}},\ \bibinfo {pages} {095020} (\bibinfo {year} {2011})},\
  \Eprint {https://arxiv.org/abs/1108.1191} {arXiv:1108.1191 [hep-ph]}
  \BibitemShut {NoStop}%
\bibitem [{\citenamefont {Banados}\ \emph {et~al.}(2006)\citenamefont
  {Banados}, \citenamefont {Miskovic},\ and\ \citenamefont
  {Theisen}}]{Banados:2006fe}%
  \BibitemOpen
  \bibfield  {author} {\bibinfo {author} {\bibfnamefont {M.}~\bibnamefont
  {Banados}}, \bibinfo {author} {\bibfnamefont {O.}~\bibnamefont {Miskovic}},\
  and\ \bibinfo {author} {\bibfnamefont {S.}~\bibnamefont {Theisen}},\
  }\bibfield  {title} {\bibinfo {title} {{Holographic currents in first order
  gravity and finite Fefferman-Graham expansions}},\ }\href
  {https://doi.org/10.1088/1126-6708/2006/06/025} {\bibfield  {journal}
  {\bibinfo  {journal} {JHEP}\ }\textbf {\bibinfo {volume} {06}},\ \bibinfo
  {pages} {025}},\ \Eprint {https://arxiv.org/abs/hep-th/0604148}
  {arXiv:hep-th/0604148} \BibitemShut {NoStop}%
\bibitem [{\citenamefont {Gubser}\ \emph {et~al.}(1998)\citenamefont {Gubser},
  \citenamefont {Klebanov},\ and\ \citenamefont {Polyakov}}]{Gubser:1998bc}%
  \BibitemOpen
  \bibfield  {author} {\bibinfo {author} {\bibfnamefont {S.~S.}\ \bibnamefont
  {Gubser}}, \bibinfo {author} {\bibfnamefont {I.~R.}\ \bibnamefont
  {Klebanov}},\ and\ \bibinfo {author} {\bibfnamefont {A.~M.}\ \bibnamefont
  {Polyakov}},\ }\bibfield  {title} {\bibinfo {title} {{Gauge theory
  correlators from noncritical string theory}},\ }\href
  {https://doi.org/10.1016/S0370-2693(98)00377-3} {\bibfield  {journal}
  {\bibinfo  {journal} {Phys. Lett. B}\ }\textbf {\bibinfo {volume} {428}},\
  \bibinfo {pages} {105} (\bibinfo {year} {1998})},\ \Eprint
  {https://arxiv.org/abs/hep-th/9802109} {arXiv:hep-th/9802109} \BibitemShut
  {NoStop}%
\bibitem [{\citenamefont {Witten}(1998)}]{Witten:1998qj}%
  \BibitemOpen
  \bibfield  {author} {\bibinfo {author} {\bibfnamefont {E.}~\bibnamefont
  {Witten}},\ }\bibfield  {title} {\bibinfo {title} {{Anti-de Sitter space and
  holography}},\ }\href {https://doi.org/10.4310/ATMP.1998.v2.n2.a2} {\bibfield
   {journal} {\bibinfo  {journal} {Adv. Theor. Math. Phys.}\ }\textbf {\bibinfo
  {volume} {2}},\ \bibinfo {pages} {253} (\bibinfo {year} {1998})},\ \Eprint
  {https://arxiv.org/abs/hep-th/9802150} {arXiv:hep-th/9802150} \BibitemShut
  {NoStop}%
\bibitem [{\citenamefont {de~Haro}\ \emph {et~al.}(2001)\citenamefont
  {de~Haro}, \citenamefont {Solodukhin},\ and\ \citenamefont
  {Skenderis}}]{deHaro:2000vlm}%
  \BibitemOpen
  \bibfield  {author} {\bibinfo {author} {\bibfnamefont {S.}~\bibnamefont
  {de~Haro}}, \bibinfo {author} {\bibfnamefont {S.~N.}\ \bibnamefont
  {Solodukhin}},\ and\ \bibinfo {author} {\bibfnamefont {K.}~\bibnamefont
  {Skenderis}},\ }\bibfield  {title} {\bibinfo {title} {{Holographic
  reconstruction of space-time and renormalization in the AdS / CFT
  correspondence}},\ }\href {https://doi.org/10.1007/s002200100381} {\bibfield
  {journal} {\bibinfo  {journal} {Commun. Math. Phys.}\ }\textbf {\bibinfo
  {volume} {217}},\ \bibinfo {pages} {595} (\bibinfo {year} {2001})},\ \Eprint
  {https://arxiv.org/abs/hep-th/0002230} {arXiv:hep-th/0002230} \BibitemShut
  {NoStop}%
\bibitem [{\citenamefont {Bianchi}\ \emph {et~al.}(2002)\citenamefont
  {Bianchi}, \citenamefont {Freedman},\ and\ \citenamefont
  {Skenderis}}]{Bianchi:2001kw}%
  \BibitemOpen
  \bibfield  {author} {\bibinfo {author} {\bibfnamefont {M.}~\bibnamefont
  {Bianchi}}, \bibinfo {author} {\bibfnamefont {D.~Z.}\ \bibnamefont
  {Freedman}},\ and\ \bibinfo {author} {\bibfnamefont {K.}~\bibnamefont
  {Skenderis}},\ }\bibfield  {title} {\bibinfo {title} {{Holographic
  renormalization}},\ }\href {https://doi.org/10.1016/S0550-3213(02)00179-7}
  {\bibfield  {journal} {\bibinfo  {journal} {Nucl. Phys. B}\ }\textbf
  {\bibinfo {volume} {631}},\ \bibinfo {pages} {159} (\bibinfo {year}
  {2002})},\ \Eprint {https://arxiv.org/abs/hep-th/0112119}
  {arXiv:hep-th/0112119} \BibitemShut {NoStop}%
\end{thebibliography}%

\onecolumngrid
\newpage
\appendix

\centerline{{\large \bf Supplemental Material}}

\vspace{0.5cm}

In the three sections of the Supplemental Material we list our conventions as well as gather some salient intermediate details that help the reader to work through the calculations.

\vspace{7mm}

\section{Conventions}

We assume our boundary field theory to be equipped with a flat Minkowski metric that is mostly plus $(\eta_{\mu\nu})=\text{diag}(-1,1,1,1)$, $\mu,\nu=0,1,2,3$ and we take $\epsilon^{0123}=+1$ and for $\epsilon^{0123z}=+1$ the Levi--Civita symbols in four and five dimensions. 

Antisymmetrization of a tensor is defined as
\begin{equation}
    X_{[\mu_1\cdots \mu_p]}=\frac{1}{p!} \sum_{\sigma\in P(\{1,\cdots,p\})} {\rm sgn}(\sigma) X_{\mu_{\sigma(1)}\ldots \mu_{\sigma(p)}}\ ,
\end{equation}
where $P(\{1,\ldots,p\})$ is the set of permutations of $\{1,\ldots,p\}$ and ${\rm sgn}(\sigma)=+1$ for even permutations and ${\rm sgn}(\sigma)=-1$ for odd permutations.

Our conventions for the Hodge dual operations are
\begin{eqnarray}
    \star_4(\dd x^{\mu_1}\wedge \cdots \wedge \dd x^{\mu_p}) &=&\frac{1}{(4-p)!}\epsilon^{\mu_1\cdots \mu_p}_{\phantom{\mu_1\cdots \mu_p}\mu_{p+1}\cdots \mu_4} \dd x^{\mu_{p+1}}\wedge \cdots \wedge \dd x^{\mu_4} \\  
    \star (e^{a_1}\wedge \cdots \wedge e^{a_p}) &=&\frac{1}{(5-p)!}\epsilon^{a_1\cdots a_p}_{\phantom{a_1\cdots a_p}a_{p+1}\cdots a_5} e^{a_{p+1}}\wedge \cdots \wedge e^{a_5}\,.
\end{eqnarray}

The four-dimensional Dirac gamma matrices are defined in the usual way, they satisfy the Clifford algebra
\begin{equation}
    \{ \gamma^\mu, \gamma^\nu\}=2\eta^{\mu\nu}\mathbb{1} \ ,
\end{equation}
giving anti-Hermitean $(\gamma^0)^\dagger=-\gamma^0$ and Hermitean $(\gamma^i)^\dagger=\gamma^i$, $i=1,2,3$. The full Dirac algebra is comprised of the identity matrix and antisymmetric products of the gamma matrices $\Gamma\in \{ \mathbb{1},\gamma^\mu, \gamma^{[\mu_1}\gamma^{\mu_2]},\gamma^{[\mu_1}\gamma^{\mu_2}\gamma^{\mu_3]},\gamma^{[\mu_1}\gamma^{\mu_2}\gamma^{\mu_3}\gamma^{\mu_4]} \}$. To each element we can associate a bilinear fermion operator $\overline{\psi}\Gamma\psi$ with $\bar\psi=\psi^\dagger i\gamma^0$, which is an antisymmetric tensor of rank zero to four.

Alternatively, one can introduce the Hermitean matrix $\gamma_5=i\gamma^0 \gamma^1 \gamma^2 \gamma^3$ and use the algebraic identities
\begin{equation}
    \gamma^{[\mu_1}\gamma^{\mu_2}\gamma^{\mu_3]}=i \epsilon^{\mu_1\mu_2\mu_3}_{\phantom{\mu_1\mu_2\mu_3}\nu}\gamma^\nu \gamma_5\ , \qquad  \gamma^{[\mu_1}\gamma^{\mu_2}\gamma^{\mu_3}\gamma^{\mu_4]}=-i\epsilon^{\mu_1\mu_2\mu_3\mu_4}\gamma_5 \ .
\end{equation}
This maps higher rank tensors formed with fermion bilinears to lower rank pseudo-tensors. A related identity also exists for the product of two gamma matrices
\begin{equation}
    \gamma^{[\mu_1}\gamma^{\mu_2]}=i \epsilon^{\mu_1\mu_2}_{\phantom{\mu_1\mu_2}\nu_1\nu_2}\gamma^{[\nu_1}\gamma^{\nu_2]} \gamma_5 \ .
\end{equation}

\section{Properties of solutions to bulk equations of motion}

The coefficients of the logarithmic terms in the expansion of the solutions to the second-order equations \eqref{eq:formexp} are $k_p^{\parallel,\perp A}=-\frac{L^2}{2}\alpha^{\parallel,\perp A} \Box c_p^{\parallel,\perp A}$ ($\Box\equiv \eta^{\mu\nu}\partial_\mu\partial_\nu$), with
\begin{equation}
(\alpha_0,\alpha_1^{\perp},\alpha_2^{\perp},\alpha_2^{\parallel},\alpha_3^{\perp},\alpha_4)^\pm= (1,1,1,-1,1,3)\ .
\end{equation}
The labels $\perp$ and $\parallel$ refer to transverse and longitudinal components, respectively. For the odd forms the longitudinal components are undetermined due to the gauge invariance of the second-order equations, they would vanish in the radial gauge $C_z^\pm=0,C_{z\mu\nu}^\pm=0$.

For the subleading coefficients
\begin{align}\label{eq:coefrel}
(j_0,j_2,j_4)^\pm & =\mp \star_4\left(\mp\frac{j_4}{3}\pm\frac{k_4}{9},j_2+k_2^\perp-k_2^\parallel,3j_0+k_0\right)^\mp \notag   \\
 (j_3^{\parallel},k_3^{\parallel },j_3^{\perp })^\pm & =\left(\pm\frac{\star_4(2j_1+k_1)^{\perp\mp}}{m_3L},  \frac{\mp 2\star_4 k_1^{\perp\mp}}{m_3L},\frac{i S_3}{4}\right) \ .  
\end{align}

\section{Details on holographic renormalization}

The action is regularized by introducing a cutoff at $z=\epsilon$. We then add a counterterm action with the fields evaluated at the cutoff that render the action finite in the limit $\epsilon\to 0$. Denoting by $h_{\mu\nu}$ the induced metric at the cutoff, the counterterm action is ($\sigma_p'=\sigma_p$ for $p$ even)
\begin{equation}
\begin{split}
    S_{ct}=&-\sum_{p=0}^4 \frac{\sigma_p'}{p!}\int_{z=\epsilon} \dd^4x \sqrt{-h}\delta_{AB}\bigg(\beta_p C_{\mu_1\cdots \mu_p}^AC^{B\,\mu_1\cdots \mu_p}\\
    &\qquad\qquad +\left(\gamma_p\log\frac{z}{L}+L\kappa_p^A \right)F^A_{\mu_1\cdots \mu_{p+1}}F^{B\,\mu_1\cdots \mu_{p+1}}\\
    &\qquad\qquad+\left(\tilde\gamma_p\log\frac{z}{L}+L\tilde\kappa_p^A \right)\partial^\rho C^B_{\rho \mu_1\cdots \mu_{p-1}}\partial_\sigma C^{A\,\sigma \mu_1\cdots \mu_{p-1}}\bigg)\\&+\frac{L^2}{\kappa^2g_\omega^2}\int_{z=\epsilon} \dd^4x \sqrt{-h}\left(-\frac{L}{4}\log\frac{z}{L}+L\kappa_\omega\right)R^a_{\ b\mu\nu}R^{b\ \mu\nu}_{\ a}\\
    &+ \frac{\alpha_1}{3!} \int_{z=\epsilon} \dd^4 x \epsilon_{AB}\bigg( \epsilon^{\mu\nu\lambda\rho} c_\mu^A \tilde C_{\nu\lambda\rho}^{B}\\
    &\qquad\qquad+\left(\frac{L^2}{2}\log \frac{z}{L}+L^2\hat\kappa_3^A\right)h^{\alpha\beta}\epsilon^{\mu\nu\lambda\rho} F_{\alpha\mu}^A \tilde F_{\beta\nu\lambda\rho}^B\bigg) \ ,
    \end{split}
\end{equation}
where $\tilde C_3^\pm=C_3^\pm-\frac{3}{2}\Omega_3$ and $A,B=\pm$.  
All indices are raised and lowered with the induced metric and $F^A_{\mu_1\cdots \mu_{p+1}}=\partial_{[\mu_1}C^A_{\mu_2\cdots \mu_{p+1}]}$, $\tilde{F}^A_{\mu_1\mu_2\mu_3}=\partial_{[\mu_1}\tilde{C}^A_{\mu_2\mu_3]}$ are the field strengths in the field theory directions. $c^A_\mu$ is the coefficient of the leading term of the series expansion in $z$ of $C_\mu^A$. The coefficients $\beta_p,\gamma_p,\tilde{\gamma}_p,\kappa_p^A, \tilde{\kappa}_p^A, \hat{\kappa}_3^A, \kappa_\omega$ are fixed by requiring that the renormalized action is finite in the limit $\epsilon \rightarrow 0$ and by imposing the fulfillment of the Hodge duality relations. The first condition gives us
\begin{subequations}
\begin{align}
(\beta_0,\beta_2,\beta_3,\beta_4) & =\frac{1}{L}\bigg(-\frac{1}{2},\frac{1}{2},1,\frac{3}{2}\bigg)\\
(\gamma_0,\gamma_1,\gamma_2,\gamma_3,\tilde{\gamma}_2,\tilde{\gamma}_4) & =-L\bigg(\frac{1}{2},\frac{1}{4},\frac{1}{6},\frac{1}{8},-1,6\bigg)\\
(\tilde \kappa_0^A,\tilde \kappa_1^A,\tilde \kappa_3^A, \kappa_4^A,\beta_1,\gamma_4,\tilde{\gamma}_0,\tilde{\gamma}_1,\tilde{\gamma}_3) & =(0,...,0) \ .      
\end{align}
\end{subequations}
The free coefficients $(\kappa_0^A,\kappa_1^A, \kappa_3^A, \tilde \kappa_4^A,\kappa_\omega, \hat \kappa_3^A)$  correspond to finite counterterms allowed by symmetries.

The expectation values of the bilinear operators are determined by the variation $\corr{\delta I_p}=\delta S_\textrm{on-shell}+\delta S_{B1}+\delta S_{B2}+\delta S_{ct}$. If $\alpha_1=0$, $\sigma_3'\neq 0$, we obtain
\begin{equation}\label{vevJ}
\!\!\!\!\! \corr{\cJ_p^\pm}=\frac{\sigma_p'}{L} \bigg(2 j_p-\frac{L^2}{2}\Box (c_p+\delta_p^{\perp}\kappa_p c_p^{\perp}+\delta_p^{\parallel} \tilde \kappa_p c_p^{\parallel})\bigg)^\pm \ .
\end{equation}
The spin current resulting from the variation of the renormalized action with respect to the spin-connection is
\begin{equation}
    \corr{\cS_3}=-\frac{L}{\kappa^2 g_\omega^2}\left(2 S_3-\frac{L^2}{2}\square \omega_3^{(0)}(1+8\kappa_\omega) \right)-\frac{i\beta}{4\kappa^2\lambda_1}\left(\corr{\cJ_3^+}_{\alpha_1}+\corr{\cJ_3^-}_{\alpha_1} \right).
\end{equation}
Note that $c_0^\perp=c_0$, $c_4^\parallel=c_4$, $c_1\to c_1^\perp$, $c_3\to c_3^\perp$ in \eqref{vevJ}. We find
\begin{equation}
(\delta_0^\perp,\delta_1^\perp,\delta_2^\perp,\delta_2^\parallel,\delta_3^\perp,\delta_4^\parallel)^\pm=(4,8,12,2,16,1) \ .
\end{equation}
Using \eqref{eq:coefrel}, one can check that the Hodge duality relations \eqref{eq:vevduality} for even forms are satisfied provided
\begin{equation}
\tilde \kappa_4^\pm=12 \kappa_0^\pm \ , \  \kappa_2^-=-\kappa_2^+\ ,\ \tilde \kappa_2^-=-\tilde \kappa_2^+ \ .
\end{equation}
For the odd forms and the spin current we fix
\begin{equation} \kappa_1^\pm=0\ , \ \kappa_3^\pm=-\frac{1}{16}\ , \ \hat{\kappa}_3^\pm=-\frac{1}{4}\ ,  \  \kappa_\omega=-\frac{1}{8} \ . 
\end{equation}
\end{document}